\documentclass[11pt]{iopart}
\usepackage{iopams}
\usepackage{graphicx}

\renewcommand{\d}[1]{\mathinner{d#1}} 
\newcommand{\fn}[2]{\mathinner{#1\mathopen{\left(#2\right)}}} 

\begin{document}

\title{The causal structure of dynamical charged black holes}

\author{Sungwook E Hong\footnote{eostm@muon.kaist.ac.kr}, Dong-il Hwang\footnote{enotsae@gmail.com}, Ewan D Stewart, Dong-han Yeom\footnote{innocent@muon.kaist.ac.kr}}
\address{Department of Physics, KAIST, Daejeon 305-701, Republic of Korea}

\begin{abstract}
We study the causal structure of dynamical charged black holes, with a sufficient number of massless fields, using numerical simulations.
Neglecting Hawking radiation, the inner horizon is a null Cauchy horizon and a curvature singularity due to mass inflation.
When we include Hawking radiation, the inner horizon becomes space-like and is separated from the Cauchy horizon, which is parallel to the out-going null direction.
Since a charged black hole must eventually transit to a neutral black hole, we studied the neutralization of the black hole and observed that the inner horizon evolves into a space-like singularity, generating a Cauchy horizon which is parallel to the in-going null direction.
Since the mass function is finite around the inner horizon, the inner horizon is regular and penetrable in a general relativistic sense.
However, since the curvature functions become trans-Planckian, we cannot say more about the region beyond the inner horizon, and it is natural to say that there is a ``physical'' space-like singularity.
However, if we assume an exponentially large number of massless scalar fields, our results can be extended beyond the inner horizon.
In this case, strong cosmic censorship and black hole complementarity can be violated.
\end{abstract}

\section{\label{sec:intro}Introduction}

\begin{figure}
\begin{center}
\includegraphics[scale=0.45]{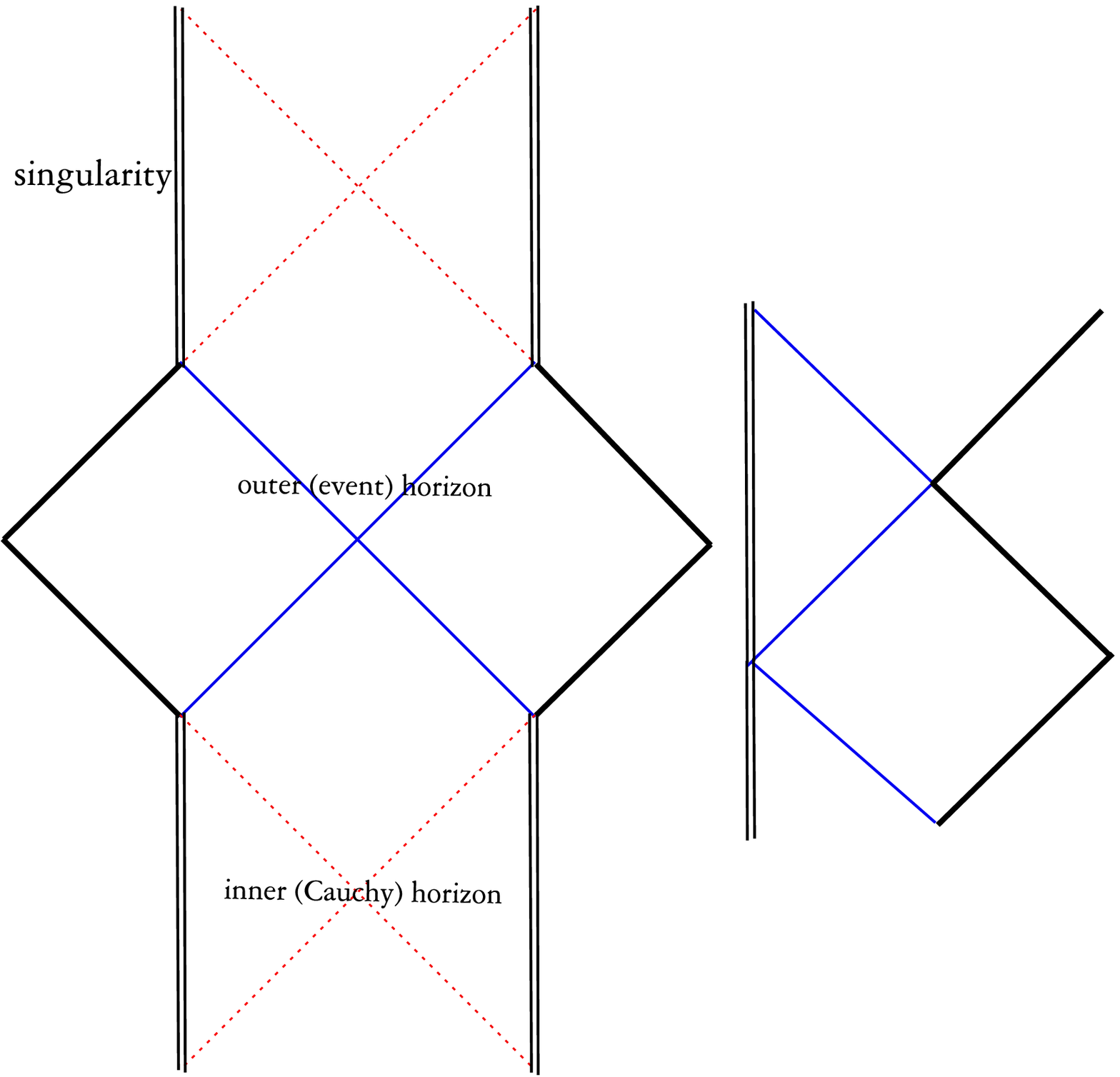}
\caption{The Penrose diagrams of static charged black holes for $M > Q$ and $M = Q$. \label{fig:charged_metric}}
\includegraphics[scale=0.6]{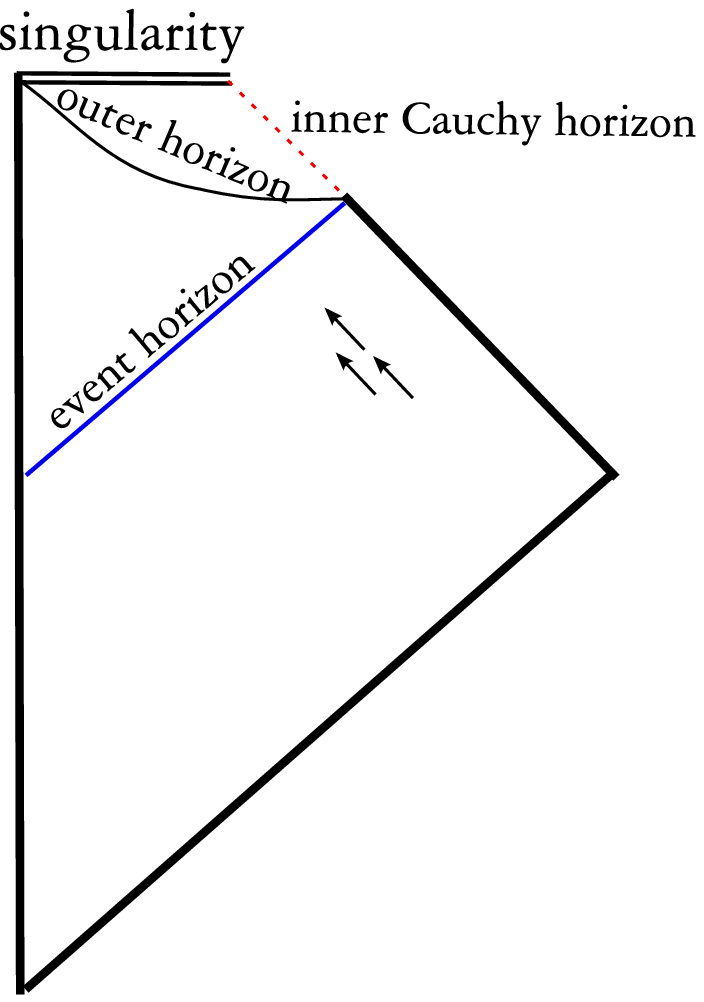}
\caption{Formation of a charged black hole with mass inflation, but without Hawking radiation. There is a null inner Cauchy horizon and a space-like singularity deep inside of the black hole. \label{fig:charged_with_mass_inflation}}
\end{center}
\end{figure}

The Reissner-Nordstrom black hole \cite{RN} has been a well-known solution since black hole physics began, see Figure~\ref{fig:charged_metric}.
However, although the geometrical structure of the static charged black hole solution was well understood \cite{Hawking:1973uf}, there still remained some interesting questions.

For example, the Reissner-Nordstrom black hole has a time-like singularity, and this may imply that there exist causally undetermined regions.
The boundary between the determined and undetermined regions is called the Cauchy horizon \cite{Wald:1984rg}.
Some authors argued that there may be local effects like infinite blue shift \cite{cc} or semi-classical effects \cite{Hiscock_etal} along the Cauchy horizon and that these effects might imply that an observer will not be able to penetrate the Cauchy horizon. This would rescue the philosophical idea known as \textit{strong cosmic censorship}, which states that no singularity is ever visible to any observer \cite{Wald:1984rg}.

In a pioneering work \cite{Poisson:1990eh}, Poisson and Israel discovered \textit{mass inflation} as an important property of the inner Cauchy horizon.
If there exists an energy flow along the inner horizon, and if an observer crosses it, he will feel infinite energy.
This means that the inner horizon is unstable and now we understand that a realistic inner horizon must be interpreted as a kind of curvature singularity.

However, there still remains an important question: \textit{what will be the back-reaction of the instability on the causal structure?}
There has been some controversy on this issue.
Some authors argued that the inner horizon\footnote{In this paper, we use the outer and inner horizons in a local sense, i.e. we will use ``apparent'' or ``trapping'' horizons \cite{local_horizon}.} is unstable and it may collapse to the singularity at the center \cite{Gnedin}.
Others said that, since the inner horizon is problematic, there should be no inner horizon in real situations due to pair-creation of charged particles \cite{FKT} (see also \cite{Ori:2006vv}).
Others argued that although the inner horizon has some problems, like a curvature singularity, it is stable and the inside structure seems to be safe, in the sense that the metric perturbation is finite and the tidal deformation is small enough \cite{Ori}\cite{Burko:2002qr}.
In other contexts, some authors argued that there are two kinds of singularities, strong and weak singularities \cite{Tipler}, and that the inner horizon becomes a weak and null singularity \cite{Burko_etal}, that is, an observer may hit the inner horizon first and then fall to the strong singularity deep inside of the black hole \cite{Bonanno:1994ma}.
However, all of these ideas were based on a charged black hole without Hawking radiation.

To figure out which opinion is correct and which is not, we need to understand the causal structure of charged black holes not only for static cases but also for \textit{dynamical} cases which include the formation and the evaporation of the black hole.
Of course, some previous authors, motivated by the information loss problem, considered the effect of Hawking radiation.
They calculated the near extreme limit of a charged black hole using some approximations \cite{near_extreme}, or the Vaidya metric \cite{Vaidya}\cite{Levin:1996qt}, but these calculations could not include the mass inflation effect.
The first successful implementation of mass inflation was in Hod and Piran \cite{HodPiran} by using numerical simulations.

Pioneering black hole simulations were done in \cite{Piran}\cite{HodPiran}\cite{SorkinPiran}\cite{OrenPiran}\cite{Hansen:2005am}(see also \cite{FKT}) whose methods we followed. In \cite{HodPiran}, the formation of a charged black hole from collapsing charged matter fields was studied, and a space-like singularity and a null inner Cauchy horizon were obtained (Figure~\ref{fig:charged_with_mass_inflation}). These results were confirmed and refined in \cite{OrenPiran}\cite{Hansen:2005am}. The first numerical simulation to include Hawking radiation for a charged black hole was done in \cite{SorkinPiran}, and it was observed that the outer and inner apparent horizons approach along the out-going null direction.

In our simulation, we assume spherical symmetry and set up equations for Maxwell and scalar fields including Hawking radiation.
After reproducing the results of previous authors, we did our own experiments about charged black holes with Hawking radiation and discharge.
Especially, we focused on the causal structure and its physical implications which were not deeply discussed by previous authors.

In this paper, we will answer the following questions:
\begin{itemize}
  \item What is the Penrose diagram for dynamical charged black holes with Hawking radiation and discharge?
  \item Is the inner horizon stable or unstable?
Is the inner horizon penetrable or not?
  \item What are the implications for cosmic censorship and black hold complementarity?
\end{itemize}
In Section \ref{sec:numerical}, we describe our model for a dynamical charged black hole.
In Section \ref{sec:res1}, we discuss the causal structure of dynamical charged black holes.
In Section \ref{sec:res2}, we discuss the properties of the inner horizon.
In Section \ref{sec:cosmic}, we discuss the implications for strong cosmic censorship and black hole complementarity.

\section{\label{sec:numerical}Model for a dynamical charged black hole}

We will briefly introduce the model used in this paper. For details, see Appendices \ref{sec:BasicSchemes}, \ref{sec:initialCond}, and \ref{sec:consistency}. In this paper, we set $G=c=k_{\mathrm{B}}=4\pi \epsilon_{0}=1$, and $m_{\mathrm{Pl}}=l_{\mathrm{Pl}}=\sqrt{\hbar}$, where $m_{\mathrm{Pl}}$ and $l_{\mathrm{Pl}}$ are the Planck mass and length.

We assume a complex massless scalar field $\phi$ which is coupled to the electromagnetic field $A_{\mu}$ \cite{Hawking:1973uf}:
\begin{eqnarray} \label{Lagrangian}
\mathcal{L} = - \left(\phi_{;a}+ieA_{a}\phi\right)g^{ab}\left(\overline{\phi}_{;b}-ieA_{b}\overline{\phi}\right)-\frac{1}{8\pi}F_{ab}F^{ab},
\end{eqnarray}
where $F_{ab}=A_{b;a}-A_{a;b}$, and $e$ is the gauge coupling.
We assume spherical symmetry
\begin{eqnarray} \label{double_null}
\d{s^{2}} = -\fn{\alpha^{2}}{u,v} \d{u} \d{v} + \fn{r^{2}}{u,v} \d{\Omega^{2}},
\end{eqnarray}
where $u$ is the retarded time, $v$ is the advanced time, and $\theta$ and $\varphi$ are angular coordinates \cite{Hamade:1995ce}.
Spherical symmetry fixes $A_{\theta} = A_{\varphi} = 0$, and we can choose the gauge $A_{v}=0$ giving $A_{\mu}=(a,0,0,0)$ \cite{OrenPiran}.

To obtain the metric, we use the semi-classical Einstein equation:
\begin{eqnarray} \label{Einstein}
G_{\mu\nu}=8\pi \left( T^{\mathrm{C}}_{\mu\nu}+\langle \hat{T}^{\mathrm{H}}_{\mu\nu} \rangle \right),
\end{eqnarray}
where the right hand side contains the energy-momentum tensor for classical collapsing fields $T^{\mathrm{C}}_{\mu\nu}$ and Hawking radiation $\langle \hat{T}^{\mathrm{H}}_{\mu\nu} \rangle$.
For the expectation value of operators to have physical meaning on a classical background, the operator $\hat{T}^{\mathrm{H}}_{\mu\nu} - \langle \hat{T}^{\mathrm{H}}_{\mu\nu} \rangle$ of all possible quantum fluctuations should be sufficiently smaller than $\langle \hat{T}^{\mathrm{H}}_{\mu\nu} \rangle$, or roughly, the normalized dispersion $[\langle (\hat{T}^{\mathrm{H}})^{2} \rangle - \langle \hat{T}^{\mathrm{H}} \rangle^{2}]/ \langle \hat{T}^{\mathrm{H}} \rangle^{2}$ should be sufficiently small.
If we have a sufficiently large number $N$ of massless fields, and if each field contributes independently to the energy-momentum tensor, then the normalized dispersion will decrease as $1/N$. Therefore, using Equation (\ref{Einstein}) is justified in the large $N$ limit.

To include the Hawking radiation in the form of the renormalized energy-momentum tensor $\langle \hat{T}^{\mathrm{H}}_{\mu\nu} \rangle$, we used $1+1$ dimensional results \cite{Davies:1976ei} divided by $4 \pi r^{2}$,
\begin{eqnarray} \label{semiclassical}
\langle \hat{T}^{\mathrm{H}}_{uu} \rangle &=& \frac{P}{4\pi r^{2} \alpha^{2}}\left(\alpha \alpha_{uu} - 2 {\alpha_{u}}^{2}\right),
\nonumber \\
\langle \hat{T}^{\mathrm{H}}_{uv} \rangle = \langle \hat{T}^{\mathrm{H}}_{vu} \rangle &=& -\frac{P}{4\pi r^{2} \alpha^{2}}\left(\alpha\alpha_{uv}-\alpha_{u}\alpha_{v}\right),
\nonumber \\
\langle \hat{T}^{\mathrm{H}}_{vv} \rangle &=& \frac{P}{4\pi r^{2} \alpha^{2}}\left(\alpha \alpha_{vv} - 2 {\alpha_{v}}^{2}\right).
\end{eqnarray}
This is a reasonable model for the spherically symmetric case \cite{SorkinPiran}. In Equation (\ref{semiclassical}), a subscript of a function denotes a partial derivative, and
\begin{eqnarray} \label{P}
P \propto Nl_{\mathrm{Pl}}^2,
\end{eqnarray}
where $N$ is the number of massless scalar fields generating the Hawking radiation and $l_{\mathrm{Pl}}$ is the Planck length.
By changing $P$, we can tune the strength of quantum effects, and by changing $N$ at fixed $P$, we can tune the Planckian cutoff.

For initial conditions which make the black hole, we take the charged field configuration at the initial surface to be
\begin{eqnarray} \label{s_initial}
\phi^{\mathrm{BH}}(u_{\mathrm{i}},v)= \frac{A}{\sqrt{4\pi}} \sin ^{2} \left( \pi \frac{v-v_{\mathrm{i}}}{v_{\mathrm{f}}-v_{\mathrm{i}}} \right) \exp \left( 2 \pi i \frac{v-v_{\mathrm{i}}}{v_{\mathrm{f}}-v_{\mathrm{i}}} \right)
\end{eqnarray}
for $v_{\mathrm{i}}\leq v \leq v_{\mathrm{f}}$ and otherwise $\phi^{\mathrm{BH}}(u_{\mathrm{i}},v)=0$, where $u_{\mathrm{i}}=0$ and $v_{\mathrm{i}}=0$ are the initial retarded and advanced time, and $v_{\mathrm{f}}=20$ is the end of the pulse in the initial surface.
Finally, after fixing some parameters, we will be left with three free parameters: the gauge coupling $e$, the strength of Hawking radiation $P$, and the amplitude of the field $A$. We determine one specific simulation by choosing values for these three parameters (see \ref{sec:initialCond}).

\section{\label{sec:res1}The causal structure of dynamical charged black holes}

\subsection{\label{sec:without}Formation of a black hole via collapsing matter fields}

The static charged black hole solution is well-known:
\begin{eqnarray} \label{charged_metric}
ds^{2} = -\left(1-\frac{2M}{r}+\frac{Q^{2}}{r^{2}}\right)dt^{2}+\left(1-\frac{2M}{r}+\frac{Q^{2}}{r^{2}}\right)^{-1}dr^{2}+r^{2}d\Omega^{2}.
\end{eqnarray}
From this metric we can draw a maximally extended causal structure giving the Penrose diagrams \cite{Hawking:1973uf} in Figure~\ref{fig:charged_metric}.
However, these diagrams are not true for dynamical situations.
For example, the initial state should be generally flat, and the final state will be flat again in many cases.

\begin{figure}
\begin{center}
\includegraphics[scale=0.3]{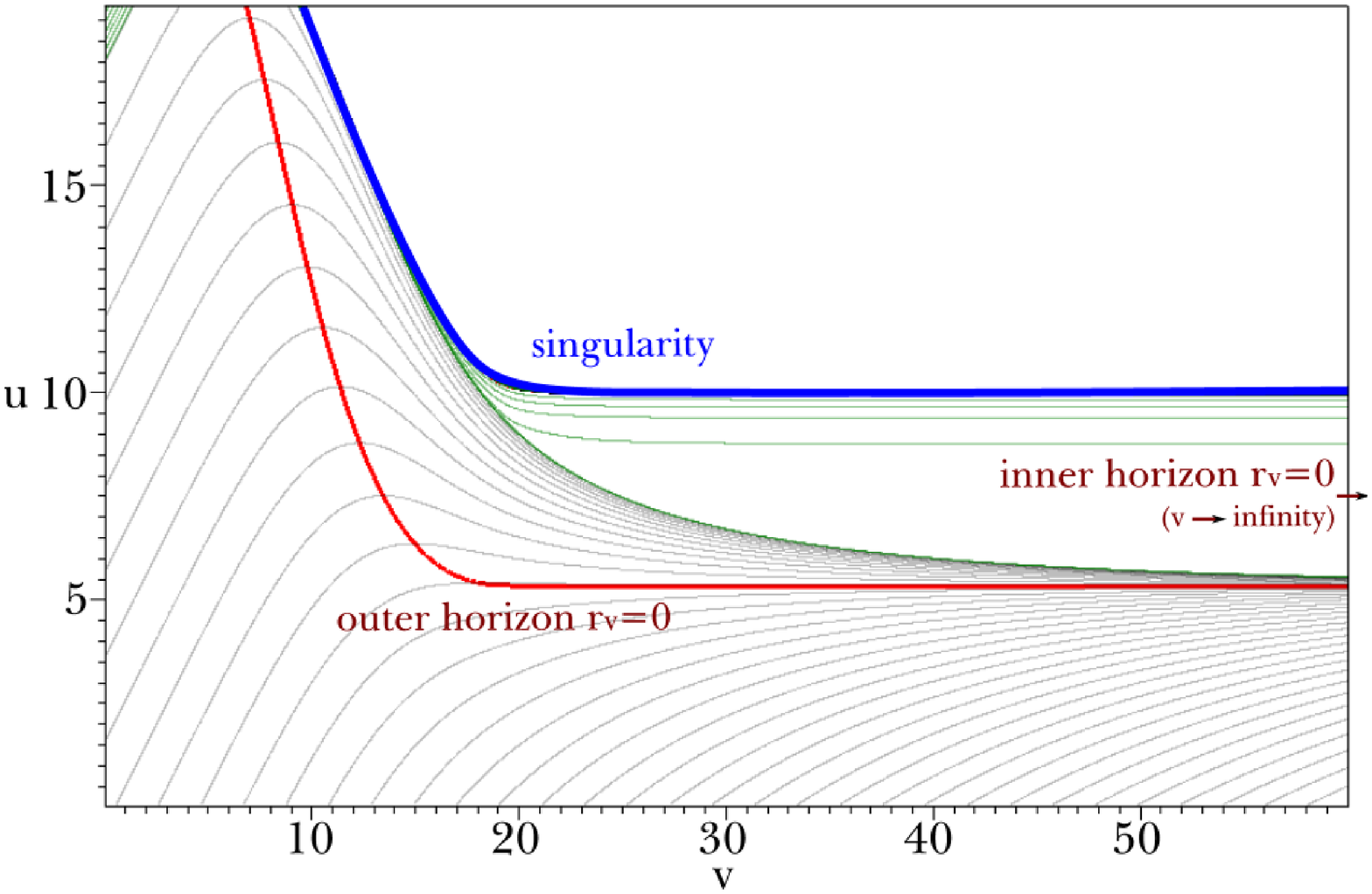}
\includegraphics[scale=0.3]{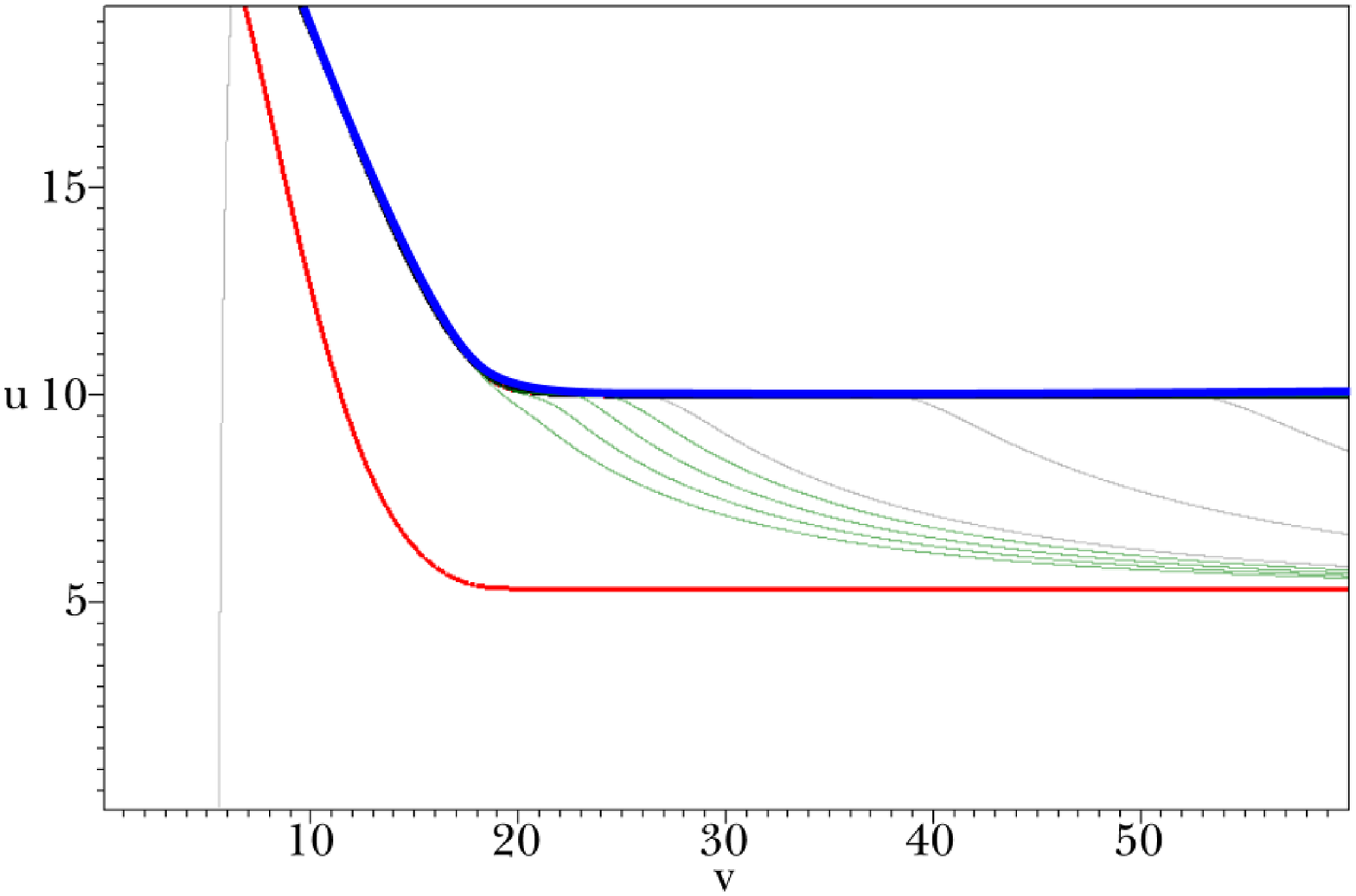}
\caption{Contour diagram of $r$ for $(e=0.1,P=0,A=0.25)$. This diagram confirms Figure~\ref{fig:charged_with_mass_inflation}. The outer horizon ($r_{v}=0$, red curve) grows in a space-like direction, and the inner horizon (also, $r_{v}=0$) would be located at the $v \rightarrow \infty$ limit. Here, spacing is $1$ for black contours and $0.1$ for green contours.\label{fig:charged_with_mass_inflation_simulation}}
\caption{The contours of the mass function for $\log |m|$ $=$ $0$, $20$, $40$, $60$, $80$, $100$, $200$, $300$.\label{fig:mass_inflation}}
\end{center}
\end{figure}

Then, what will happen if we collapse some charged matter in flat space-time?
One simple guess is the growth of outer and inner horizons, and the evolution of a time-like singularity \cite{Vaidya}.
However, because of mass inflation, which makes the inner horizon singular, it was guessed that there is a space-like singularity deep inside the black hole and a null inner horizon \cite{Ori}\cite{Bonanno:1994ma} as in Figure~\ref{fig:charged_with_mass_inflation}.
This idea was confirmed by some numerical simulations \cite{HodPiran}\cite{OrenPiran}.
Also, we reproduced the same result in Figure~\ref{fig:charged_with_mass_inflation_simulation}.

To check whether the inner horizon becomes singular or not, we need to see the behavior of the mass function.
The mass function is defined by \cite{Waugh:1986jh}
\begin{eqnarray} \label{mass}
m(u,v) \equiv \frac{r}{2}\left( 1+\frac{q^{2}}{r^{2}}+4\frac{r_{u}r_{v}}{\alpha^{2}} \right),
\end{eqnarray}
where $q(u,v) \equiv 2r^{2} a_{v}/\alpha^{2}$ is the charge (see \ref{sec:BasicSchemes}).
Figure~\ref{fig:mass_inflation} shows the behavior of the mass function. It blows up exponentially as $v \rightarrow \infty$. This is a typical signature of mass inflation, $m \sim \exp(\kappa_{\mathrm{i}}v)$, where $\kappa_{\mathrm{i}}$ is the surface gravity of the inner horizon \cite{Poisson:1990eh}.
Then, some scalar quantities such as the Ricci scalar and the Kretschmann scalar ($K \equiv R_{\alpha \beta \gamma \delta}R^{\alpha \beta \gamma \delta}\sim m^{2}$) become infinite as $v \rightarrow \infty$ \cite{Ori}.
Since we did not consider quantum effects, the Planckian cutoff of the scalar curvatures can be regarded as $\infty$.
Thus, the curvature singularity occurs at the $v \rightarrow \infty$ limit.

Note that there is a transition region between the black curves and the green curves in Figure~\ref{fig:charged_with_mass_inflation_simulation} where the decrease of the radial function begins to vary slowly. From Figure~\ref{fig:mass_inflation}, we can see that mass inflation begins around there. This region corresponds to $r \simeq M - \sqrt{M^{2}-Q^{2}}$, i.e. the inner horizon of the Reissner-Nordstrom metric, although it is not a real horizon. Its physical meaning and more detailed analysis will be included in a future work \cite{HSY}.

\subsection{\label{sec:quantum}Evaporation and discharge}

The evolution of a charged black hole is determined by two quantum effects: Hawking radiation and charge pair creation.

In a Reissner-Nordstrom metric, the rate of change of the mass depends on the Hawking temperature. From the Stefan-Boltzmann law,
\begin{eqnarray}
\frac{dM}{dt} = 4 \pi \sigma r_{+}^{2} T_{H}^{4} = \frac{1}{240\pi}\frac{(M^{2}-Q^{2})^{2}}{r_{+}^{6}},
\end{eqnarray}
where $\sigma=\pi^{2}/60$ is the Stefan-Boltzmann constant,
\begin{eqnarray}
r_{+} = M + \sqrt{M^{2}-Q^{2}}
\end{eqnarray}
is the outer horizon, and
\begin{eqnarray}
T_{H} = \frac{1}{2\pi} \frac{\sqrt{M^{2}-Q^{2}}}{r_{+}^{2}}
\end{eqnarray}
is the Hawking temperature.

The pair creation rate is
\begin{eqnarray}
\Gamma \simeq \frac{e^{2} E^{2}}{4 \pi^{3}} e^{-\frac{E_{c}}{E}}
\end{eqnarray}
where
\begin{eqnarray}
E_{c} = \frac{\pi m_{e}^{2}}{e}
\end{eqnarray}
is the critical electric field, $m_{e}$ is the electron mass, and $e$ is the gauge coupling \cite{Dunne:2004nc}. The electric field near the horizon is
\begin{eqnarray}
E_{+} = \frac{Q}{r_{+}^{2}},
\end{eqnarray}
and the pair creation rate near the horizon is $\Gamma_{+} \equiv \Gamma(E_{+})$. Therefore, the rate of change of the charge becomes approximately
\begin{eqnarray}
\frac{dQ}{dt} \sim \Gamma_{+} \mathcal{V} \sim \Gamma_{+} r_{+}^{3} \sim \frac{e^{2}Q^{2}}{r_{+}}e^{-\frac{E_{c}}{E_{+}}},
\end{eqnarray}
where $\mathcal{V}$ is the proper volume for pair creation \cite{SorkinPiran}.

\begin{figure}
\begin{center}
\includegraphics[scale=0.6]{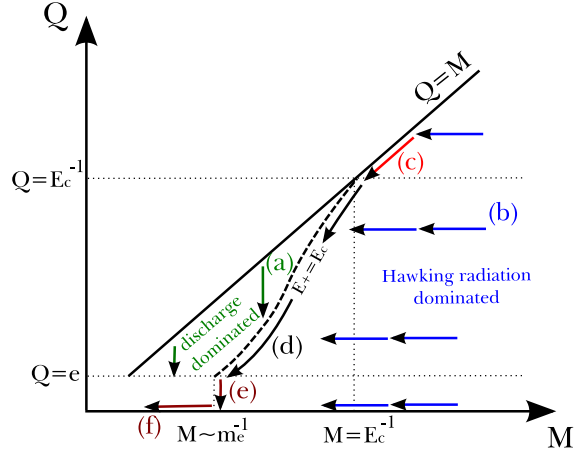}
\caption{\label{fig:M_Q}Evolution of charged black holes. (a) If $E_{+}\gg E_{c}$, discharge dominates and the black hole approaches $E_{+} = E_{c}$. (b) If $E_{+}\ll E_{c}$, discharge is exponentially suppressed. If $Q/M \ll 1$, Hawking radiation dominates and the mass will decrease via Hawking radiation. (c) If $Q \gg E^{-1}_{c}$, the black hole will approach and follow the extreme black hole (line $Q=M$). (d) If $Q \ll E^{-1}_{c}$, the black hole will approach and follow $E_{+} = E_{c}$. (e) When the charge reduces to the last few quanta, $Q \sim e$, then $T_{H}\sim M^{-1}\sim m_{e}$ on the $E_{+}=E_{c}$ track, and the black hole will emit its final quanta of charge via Hawking radiation. (f) The final neutral black hole will lose mass via Hawking radiation.}
\end{center}
\end{figure}

Using these results, we can draw a schematic diagram for the evolution of a charged black hole as in Figure~\ref{fig:M_Q}.

\begin{itemize}

\item[(a)] If $E_{+}\gg E_{c}$, then
\begin{eqnarray}
\frac{1}{M}\frac{dM}{dt} \sim \frac{1}{M}\frac{(M^{2}-Q^{2})^{2}}{r_{+}^{6}} \ll \frac{e^{2}Q}{r_{+}} \sim \frac{1}{Q}\frac{dQ}{dt},
\end{eqnarray}
and hence pair-creation dominates Hawking radiation. Then the black hole will emit charge to approach the track $E_{+} = E_{c}$:
\begin{eqnarray} \label{track}
E_{c}M = \frac{\sqrt{E_{c}Q}}{2} \left( 1+ E_{c}Q \right) \qquad \qquad (E_{c}Q\leq1).
\end{eqnarray}

\item[(b)] If $E_{+}\ll E_{c}$, discharge is exponentially suppressed and, if $Q/M \ll 1$, the black hole will lose mass via Hawking radiation.

\item[(c)] If $Q \gg E^{-1}_{c}$, then $E_{+} \ll E_{c}$ still holds for the extreme black hole.\footnote{$E_{c}^{-1} = e/\pi m_{e}^{2} \sim 10^{44} m_{\mathrm{Pl}} \sim 10^{6} M_{\mathrm{Sun}}$. To approach extremality, $Q \gg 10^{6}M_{\mathrm{Sun}}$ is required.} Therefore, the black hole will approach an extreme black hole and the Hawking radiation also becomes suppressed. The black hole will then emit mass and charge maintaining extremality.

\item[(d)] If $Q \ll E^{-1}_{c}$, the black hole will approach the $E_{+} = E_{c}$ track. The black hole will then start to emit charge and follow the track.

\item[(e)] When the charge reduces to a few quanta, $Q \sim e$, then $M \sim 1/m_{e}$ on the $E_{+}=E_{c}$ track, and the Hawking temperature becomes of the order of the electron mass, i.e. $T_{H} \sim M^{-1} \sim m_{e}$. Therefore, Hawking radiation will emit the final quanta of charge and the black hole will be neutralized \cite{Gibbons:1975kk}.

\item[(f)] After the black hole is neutralized, it will lose mass via Hawking radiation.

\end{itemize}

\begin{figure}
\begin{center}
\includegraphics[scale=0.4]{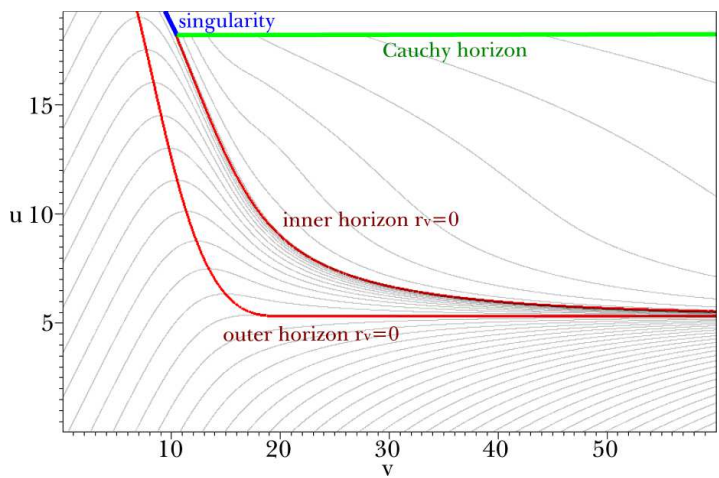}
\caption{Contour diagram of $r$ for $(e=0.1,P=0.1,A=0.25)$. Red curves show $r_{v} = 0$ horizons. \label{fig:charged_HR_G}}
\end{center}
\end{figure}

\begin{figure}
\begin{center}
\includegraphics[scale=0.3]{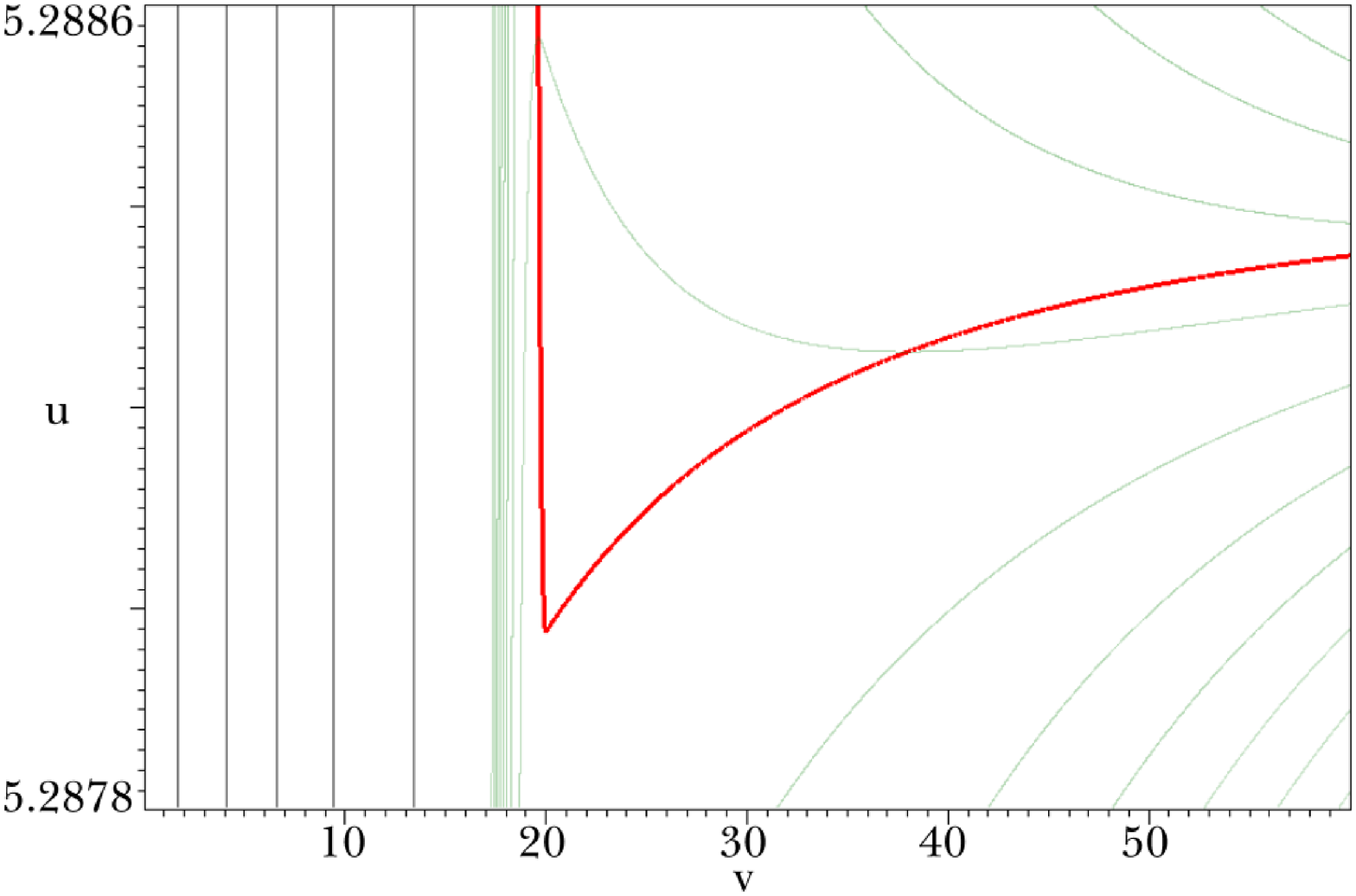}
\includegraphics[scale=0.7]{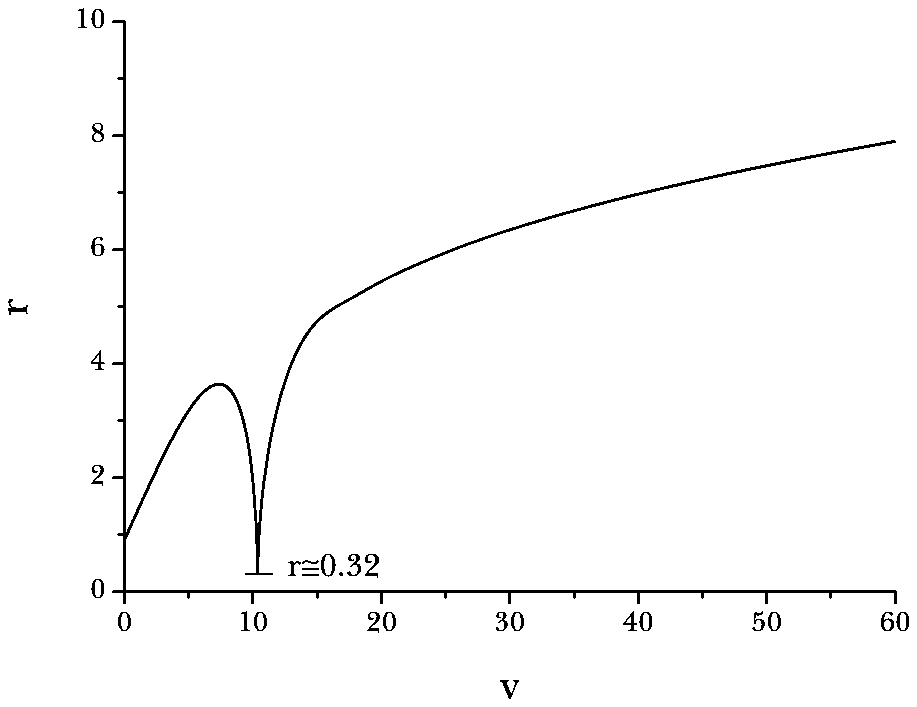}
\caption{Detailed plot near the outer horizon of Figure~\ref{fig:charged_HR_G}. This shows that the outer horizon (red curve) bends in a time-like direction. Here, spacing is $1$ for black contours and $0.002$ for green contours.\label{fig:charged_HR_R_narrow}}
\caption{$r$ function along the green cutoff line of Figure~\ref{fig:charged_HR_G}, $u \simeq 18.18$. The sharp point has $r = \sqrt{P} \simeq 0.32$, which is the central singularity. This confirms that the green cutoff line of Figure~\ref{fig:charged_HR_G} is a Cauchy horizon. \label{fig:Cauchy_proof}}
\end{center}
\end{figure}
\begin{figure}
\begin{center}
\includegraphics[scale=0.55]{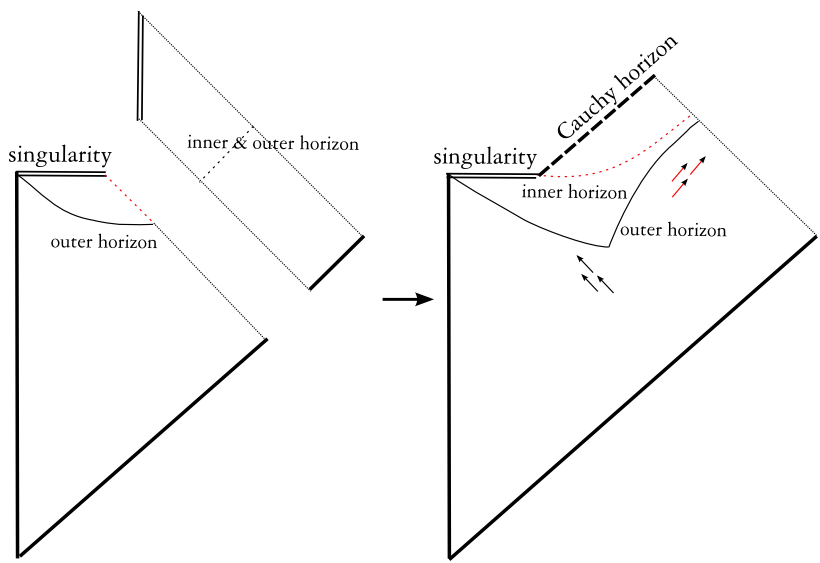}
\caption{The Penrose diagram for Figure~\ref{fig:charged_HR_G}. \label{fig:charged_HR}}
\end{center}
\end{figure}

Thus, in terms of causal structures, we need to understand two important stages:
\begin{itemize}
  \item a non-extreme black hole evolves toward an extreme black hole via Hawking radiation,
  \item the near extreme black hole evolves to a neutral black hole via discharge.
\end{itemize}
We will observe these two stages independently.

\subsubsection{\label{sec:with}From a non-extreme black hole to an extreme black hole via Hawking radiation}

In this section, we investigate a charged black hole with Hawking radiation using the parameters $(e=0.1,P=0.1,A=0.25)$.
Figure~\ref{fig:charged_HR_G} and \ref{fig:charged_HR_R_narrow} show contour diagrams of $r$ as well as the outer and inner horizons, which are defined by $r_{v}=0$. One can see that the inner horizon is space-like and approaches the out-going null direction as time $v$ increases. The space-like inner horizon is different from Section \ref{sec:without}, but agrees with \cite{SorkinPiran}.

In Figure~\ref{fig:charged_HR_G}, $r$ increases beyond the inner horizon. This means that the $r_{u}=0$ contour is almost the same as the inner horizon $r_{v}=0$. If we change the mass $M$ and the charge $Q$, the existence and properties of the $r_{u}=0$ contour and the qualitative behavior of the radial function can change. This behavior will be discussed in future work \cite{HHY}. However, the $r_{v}=0$ horizon structures are qualitatively the same.

One interesting feature is the green null cutoff line along $u \simeq 18.18$ of Figure~\ref{fig:charged_HR_G}.
We plotted the $r$ function along this line in Figure~\ref{fig:Cauchy_proof}, and it shows a sharp point at $r = \sqrt{P} \simeq 0.32$, i.e. the central singularity.
Although this cutoff line itself is regular, the region beyond cannot be calculated because of the singularity.
This null section grows from the singularity, and we can interpret the null cutoff line as the Cauchy horizon by definition.
Note that the inner horizon is separated from the Cauchy horizon.

When the mass of a black hole decrease via Hawking radiation, the inner horizon is space-like and the outer horizon is time-like \cite{local_horizon}.
Therefore, to connect the non-extreme Penrose diagram (Figure~\ref{fig:charged_with_mass_inflation}) to the extreme one (right of Figure~\ref{fig:charged_metric}), the only possible way is the time-like bending of the outer horizon, and the space-like bending of the inner horizon \cite{near_extreme}. Therefore, this logical expectation is consistent with our numerical simulations.

Finally, we can draw a schematic diagram in Figure~\ref{fig:charged_HR}.\footnote{Some previous authors pasted a near extreme black hole (left of Figure~\ref{fig:charged_metric}) to an extreme black hole (right of Figure~\ref{fig:charged_metric}) \cite{near_extreme}.}

\begin{figure}
\begin{center}
\includegraphics[scale=0.75]{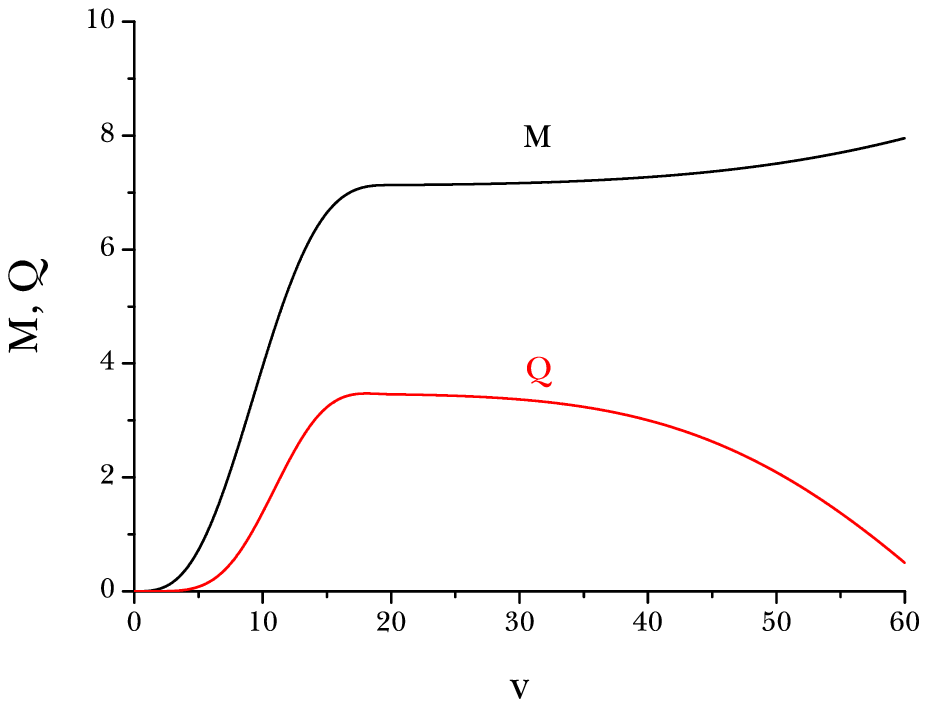}
\includegraphics[scale=0.7]{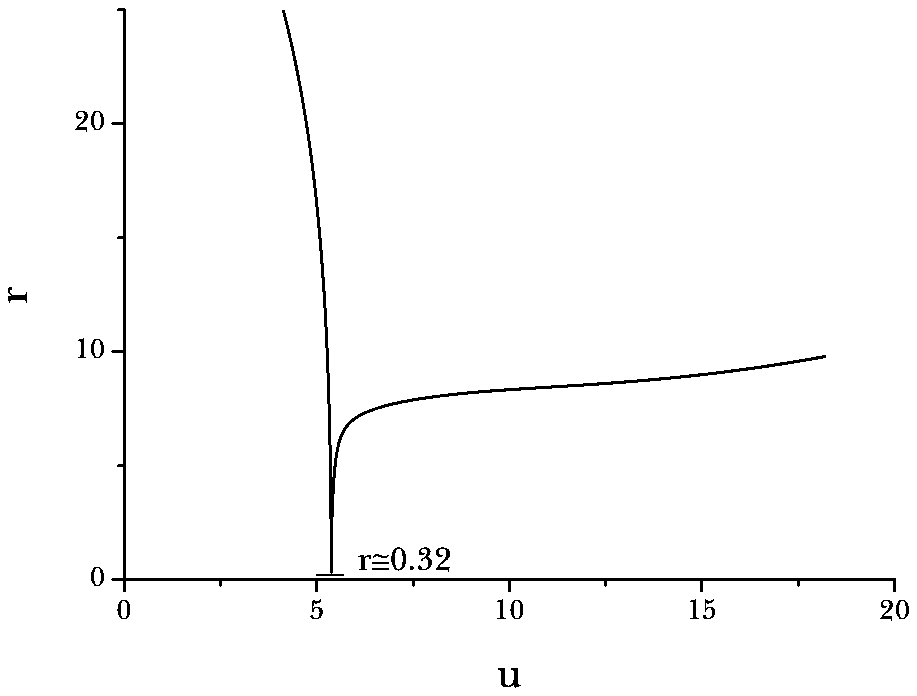}
\caption{Asymptotically calculated charge and mass during the neutralization. \label{fig:charged_neutral_Q_M}}
\caption{$r$ function along the green cutoff line of Figure~\ref{fig:charged_neutral_contour}, $v \simeq 58.11$. This confirms that the second green cutoff line is also a Cauchy horizon. \label{fig:charged_neutral_Cauchy}}
\end{center}
\end{figure}

\subsubsection{\label{sec:charged-neutral}From an extreme black hole to a neutral black hole via discharge}

In this section, we investigate a charged black hole with discharge. A transition from a charged black hole to a neutral black hole can be simulated by supplying opposite charged matter to the black hole:
\begin{eqnarray}
\phi^{\mathrm{dis}}(u_{\mathrm{i}},v)= \frac{A^{\mathrm{dis}}}{\sqrt{4\pi}} \sin ^{2} \left( \pi \frac{v-v^{\mathrm{dis}}_{\mathrm{i}}}{v^{\mathrm{dis}}_{\mathrm{f}}-v^{\mathrm{dis}}_{\mathrm{i}}} \right) \exp \left( - 2 \pi i \frac{v-v^{\mathrm{dis}}_{\mathrm{i}}}{v^{\mathrm{dis}}_{\mathrm{f}}-v^{\mathrm{dis}}_{\mathrm{i}}} \right)
\end{eqnarray}
for $v^{\mathrm{dis}}_{\mathrm{i}}\leq v \leq v^{\mathrm{dis}}_{\mathrm{f}}$ and otherwise $\phi^{\mathrm{dis}}(u_{\mathrm{i}},v)=0$, where $A^{\mathrm{dis}}=0.15$, $v^{\mathrm{dis}}_{\mathrm{i}}=0$ and $v^{\mathrm{dis}}_{\mathrm{f}}=120$. Then the total scalar field initial conditions are $\phi_{\mathrm{i}}=\phi^{\mathrm{BH}}_{\mathrm{i}}+\phi^{\mathrm{dis}}_{\mathrm{i}}$.
As we slowly add this matter, the total charge decreases greatly, but the mass increases slightly, see Figure~\ref{fig:charged_neutral_Q_M}. We may regard this experiment as a good toy model for a general charged-neutral transition.

\begin{figure}
\begin{center}
\includegraphics[scale=0.4]{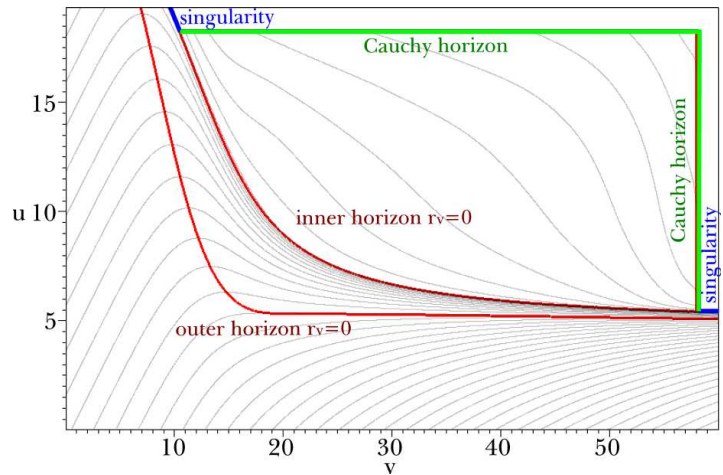}
\caption{Contour diagram of $r$. Note that, compared with Figure~\ref{fig:charged_HR_G}, there is another Cauchy horizon along the $u$ direction. \label{fig:charged_neutral_contour}}
\end{center}
\end{figure}
\begin{figure}
\begin{center}
\includegraphics[scale=0.55]{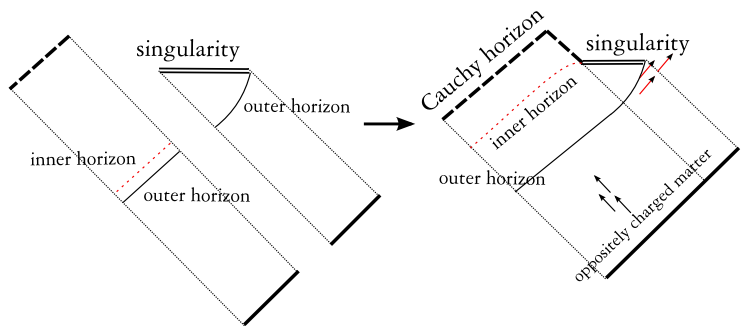}
\caption{The Penrose diagram for a transition from a near extreme charged black hole to a neutral black hole. \label{fig:charged_neutral}}
\end{center}
\end{figure}

\begin{figure}
\begin{center}
\includegraphics[scale=0.5]{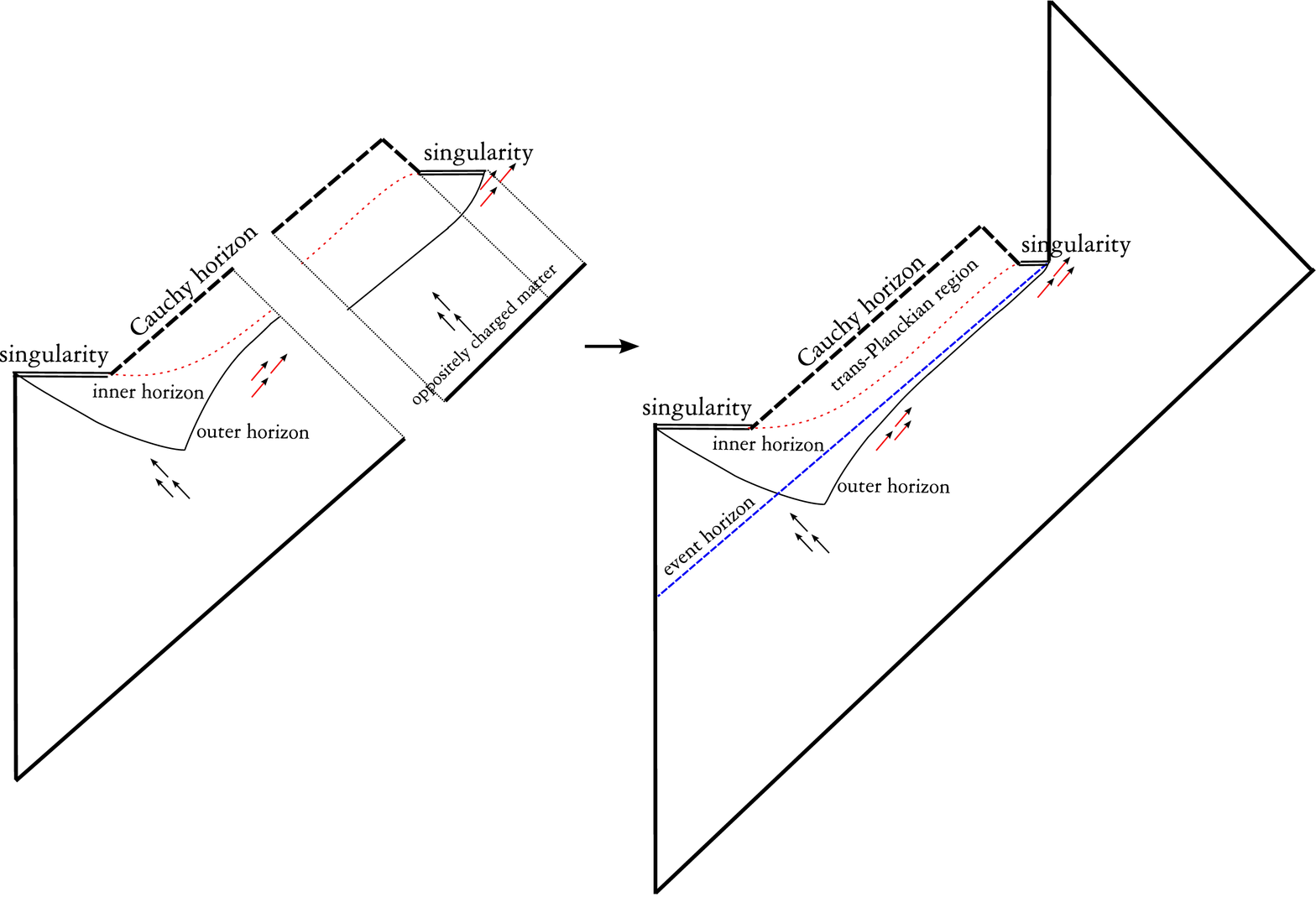}
\caption{The causal structure of dynamical charged black holes. \label{fig:charged}}
\end{center}
\end{figure}

In Figure~\ref{fig:charged_neutral_contour}, as the charge of the black hole decreases, there arises another null cutoff line in the $u$ direction.
As before, using Figure~\ref{fig:charged_neutral_Cauchy}, one can see that this cutoff line is a Cauchy horizon.

During the neutralization, the charged black hole evolves into a neutral black hole (left of Figure~\ref{fig:charged_neutral}).
Therefore, the region beyond the inner horizon should disappear, and a space-like singularity should appear. In our simulation, the inner horizon evolved into the space-like singularity, and a Cauchy horizon formed. Finally, we can draw the causal diagram for this process (right of Figure~\ref{fig:charged_neutral}). Our results are consistent with previous work \cite{Levin:1996qt}.

\subsection{\label{sec:the_causal}Conclusion}

Now we can draw the causal structure of dynamical charged black holes by combining Figures~\ref{fig:charged_HR} and \ref{fig:charged_neutral} to obtain Figure~\ref{fig:charged}. One can see the outer and inner horizons, as well as the Cauchy horizons.

\section{\label{sec:res2}Properties of inner horizons}

\subsection{\label{sec:penetrability}Penetrability of inner horizons}

The black hole was formed by the charged matter pulse $\phi^{\mathrm{BH}}$, and to test penetrability, a second neutral pulse $\phi^{\mathrm{pen}}$ was added to the initial conditions:
\begin{eqnarray}
\phi^{\mathrm{pen}}(u_{\mathrm{i}},v)= \frac{A^{\mathrm{pen}}}{\sqrt{4\pi}} \sin ^{2} \left( \pi \frac{v-v^{\mathrm{pen}}_{\mathrm{i}}}{v^{\mathrm{pen}}_{\mathrm{f}}-v^{\mathrm{pen}}_{\mathrm{i}}} \right)
\end{eqnarray}
for $v^{\mathrm{pen}}_{\mathrm{i}}\leq v \leq v^{\mathrm{pen}}_{\mathrm{f}}$ and otherwise $\phi^{\mathrm{pen}}(u_{\mathrm{i}},v)=0$,
where $A^{\mathrm{pen}}=0.01$, $v^{\mathrm{pen}}_{\mathrm{i}}=30$ and $v^{\mathrm{pen}}_{\mathrm{f}}=40$. This simulation is the same as Figure~\ref{fig:charged_HR_G} between $v=0$ and $v=30$, but modified from $v=30$.
The scalar field has initial conditions $\phi^{\mathrm{BH}}_{\mathrm{i}}+\phi^{\mathrm{pen}}_{\mathrm{i}}$.
Note that the penetrating pulse has energy $\propto (A^{\mathrm{pen}})^{2} \sim 10^{-4}$ while the energy of the first pulse is $\propto A^{2} \sim 0.06$, and hence the perturbed energy is approximately $1/100$ of the total mass.
Since the equation for the scalar field, Equation~\ref{S1}, is linear in the field and its derivatives, and the energy of the penetrating pulse is sufficiently smaller than the energy of the initial pulse, we can trace the effect of $\phi^{\mathrm{BH}}_{\mathrm{i}}$ and $\phi^{\mathrm{pen}}_{\mathrm{i}}$ independently, up to gravitational back reaction.

\begin{figure}
\begin{center}
\includegraphics[scale=0.3]{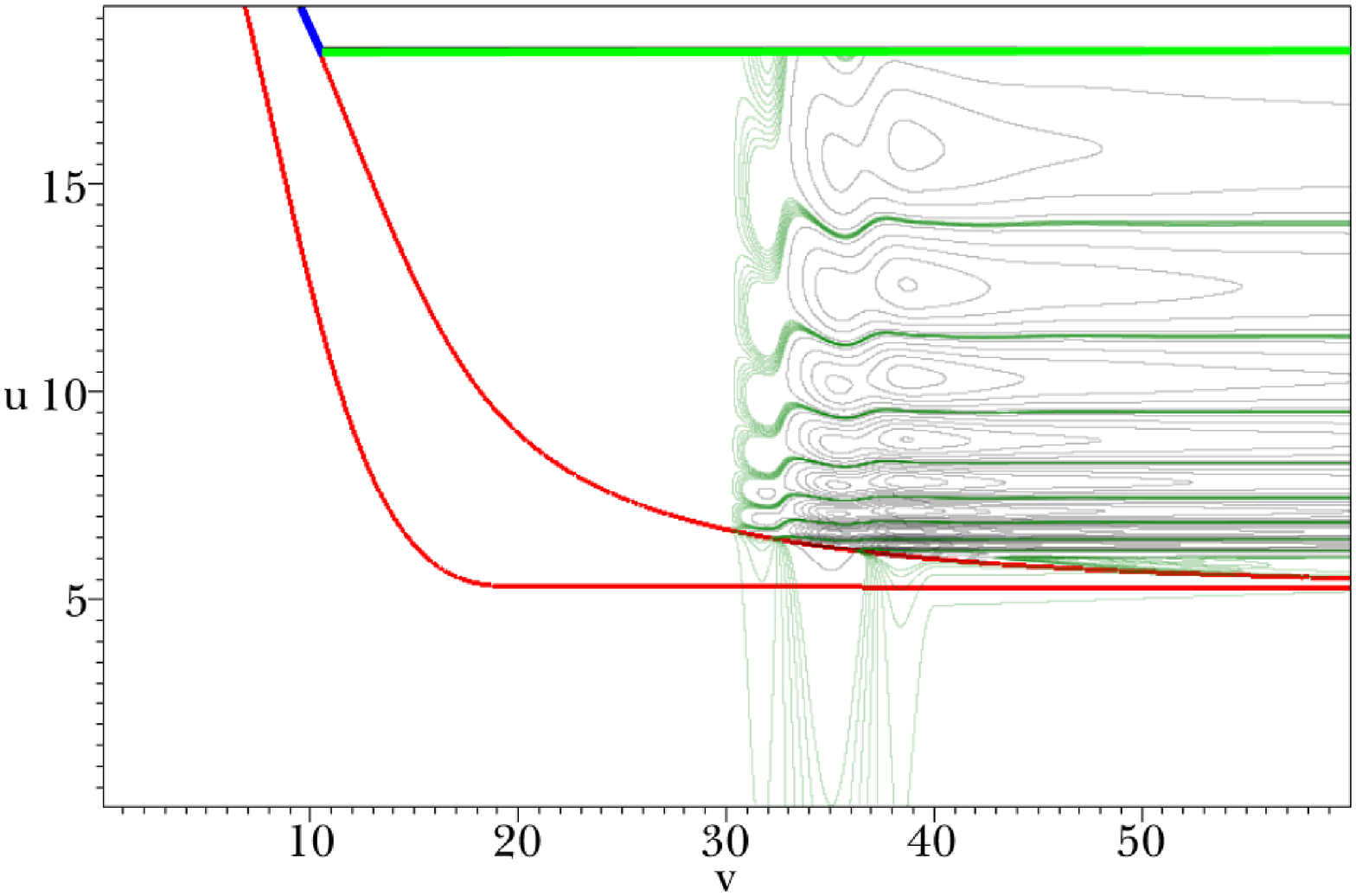}
\includegraphics[scale=0.3]{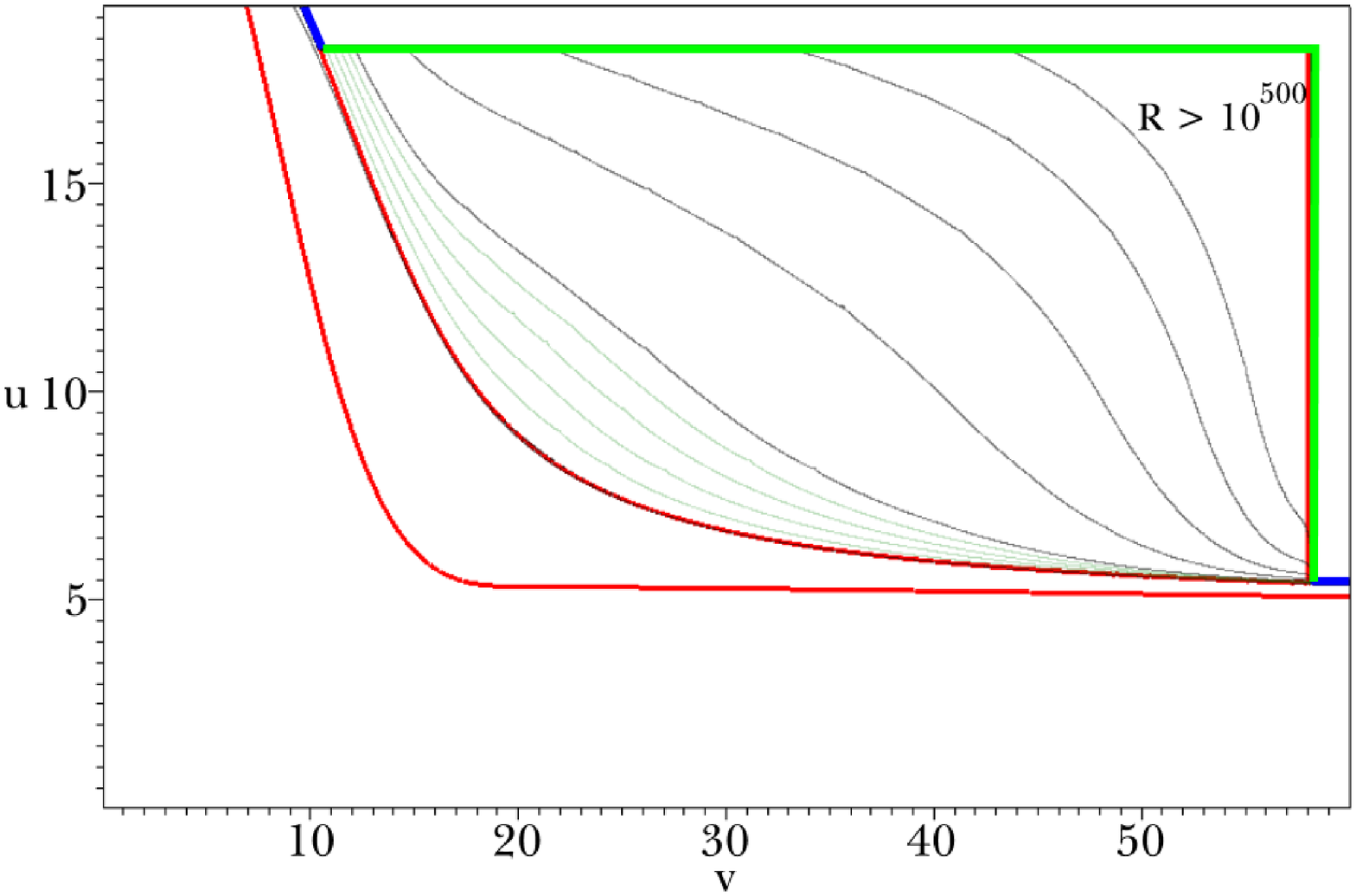}
\caption{Contour diagram of $\sqrt{4\pi} \Re(\phi^{\mathrm{pen}})$. It is affected by, but penetrates, the inner horizon. Here, spacing is $0.06457$ for black contours and $0.002$ for green contours.\label{fig:s1_real}}
\caption{The contours of the Ricci scalar for $\log |R|$ $=$ $0$, $20$, $40$, $60$, $80$, $100$, $200$, $300$, $400$, $500$.\label{fig:Ricci}}
\end{center}
\end{figure}

Figure~\ref{fig:s1_real} shows the behavior of each pulse.
In this diagram, one can see that the inner horizon has a barrier property; field configurations change drastically near the inner horizon.
However, the inner horizon is penetrable; fields penetrate the inner horizon, and flow beyond it.

\subsection{\label{sec:analysis}Mass inflation and the trans-Planckian problem}


Now we observe the curvature function for Figure~\ref{fig:charged_neutral_contour}.
Figure~\ref{fig:Ricci} shows the contours of the Ricci scalar $R$.
The Ricci scalar starts to blow up exponentially near the inner horizon, which is a typical signature of mass inflation \cite{HodPiran}\cite{OrenPiran}, but it does not diverge except at the central singularity $r=\sqrt{P}$ and it is still finite even on the Cauchy horizon.

However, since we consider quantum effects, we should also consider the Planck scale.
As the metric function $\alpha$ in Equation (\ref{double_null}) decreases exponentially beyond the inner horizon, the mass function ($\sim~1/\alpha^{2}$), the Ricci scalar ($\sim~1/\alpha^{2}$), and the Kretschmann scalar ($\sim~1/\alpha^{4}$) blow up exponentially.
When they become greater than the Planckian cutoff, the classical picture will break down.

However, from Equation~\ref{P}, the Planckian curvature cutoff, $R_{\mathrm{cutoff}} = l_{\mathrm{Pl}}^{-2} \propto N/P$, is not defined until one has defined the number of massless fields $N$. So, if we assume sufficiently large $N$, we can extend the classical picture to arbitrary large $R$. The required $N$ is of the order of the curvature $R$ beyond the inner horizon, since $R < R_{\mathrm{cutoff}} \propto N/P$ is required. The characteristic scale of $R$ is of the order of the mass function $M$, which is of the order of the mass inflation factor $\exp \kappa_{\mathrm{i}}(u+v)$ \cite{Poisson:1990eh}, where $\kappa_{\mathrm{i}}$ is the surface gravity of the inner horizon. $\kappa_{\mathrm{i}}$ is of the order of $1/M$ and $u$ and $v$ are of the order of the lifetime $M^{3}$ and hence the required $N$ is of the order of $\exp M^{2}$.

\begin{figure}
\begin{center}
\includegraphics[scale=0.5]{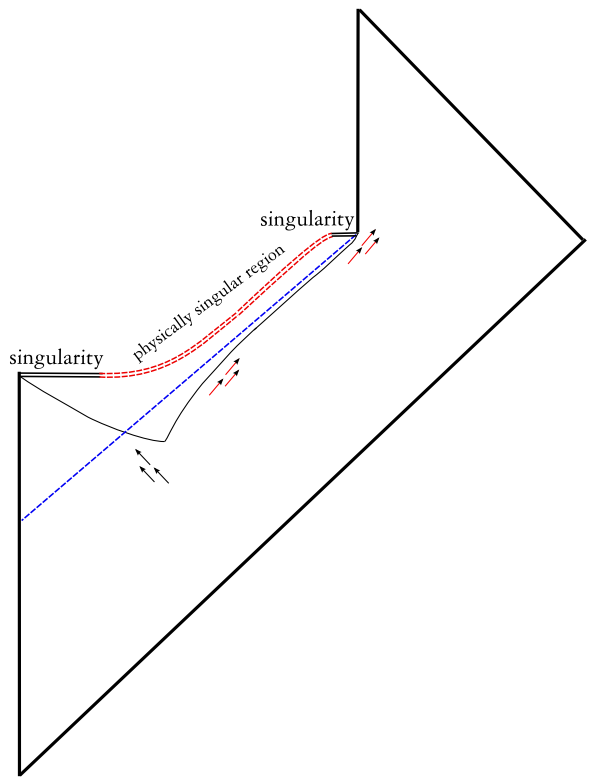}
\caption{\label{fig:charged_reasonable}The causal structure of dynamical charged black holes for a small number of massless fields. It is natural to think that there is a kind of physically singular region around the inner horizon. The region (red dashed curve) is space-like and has non-zero radius, while the central singularity has zero radius.}
\end{center}
\end{figure}

If one can assume such a large $N$, the whole region in Figure~\ref{fig:charged} becomes reliable except for the central singularity. The authors think that this would have important implications for cosmic censorship and black hole complementarity, see Section \ref{sec:cosmic}. However, if $N$ is bounded in principle or such a large $N$ spoils some assumptions of semiclassical gravity, we cannot avoid the trans-Planckian curvature problem. In this case, we should regard the trans-Plankian inner horizon as a kind of singularity (Figure~\ref{fig:charged_reasonable}).\footnote{This is quite similar to some classical dilaton black hole models \cite{Garfinkle:1990qj}, in which the inner horizon is a physical singularity with non-zero radius, and the black hole approaches a stable extreme black hole. However, there are some differences: as discussed above, the inner horizon in our model is regular in a general relativistic sense, in that it is geodesically complete and penetrable.}

\section{\label{sec:cosmic}Cosmic censorship and black hole complementarity}

Some authors have suggested that weak cosmic censorship, which states that all singularities are hidden within black holes, can be violated in charged black holes \cite{Vaidya}. However, those authors used the Vaidya metric, in which the mass and charge functions are put in by hand, which can be unphysical. Weak cosmic censorship was not violated in our simulations (see also \cite{OrenPiran}).
Whether strong cosmic censorship holds is less clear. Without Hawking radiation, the mass function diverges at the inner Cauchy horizon, and this prevents the violation of strong cosmic censorship. But if we include Hawking radiation, the mass function is finite at the Cauchy horizon, and if we assume sufficiently large $N$, all curvatures can be less than the Planckian cutoff, and hence strong cosmic censorship is violated.

Finally, using our results, we investigate black hole complementarity \cite{susskind}\cite{complementarity}.
According to black hole complementarity, after the information retention time (when the area of the black hole decreases to half its initial value), an observer outside a black hole can see the information of the in-falling matter in the Hawking radiation.
However, since the free-falling information is not affected by the Hawking radiation, two copies of the information exist, which violates the no cloning theorem of quantum mechanics. Here, black hole complementarity argues that this is not a problem since the two copies cannot be observed by a single observer. To check whether one observer can see both copies of the information or not, one may consider a gedanken experiment.
One observer free-falls into the black hole and sends his information $A$ in the out-going direction right after he crosses the event horizon.
After the information retention time, a second observer sees information $B$ in the Hawking radiation which includes $A$, and then goes into the black hole to get the information $A$. If this is possible, the second observer sees both $A$ and $B$ violating the no cloning theorem. However, in a neutral black hole, to deliver the information to the second observer before it collapses to the singularity, the first observer should send the signal with energy $\Delta E \sim \exp{M^{2}}$, which is greater than the black hole mass itself, and hence this process is impossible. This argument also holds for charged black holes with the causal structure in Figure~\ref{fig:charged_reasonable}.

\begin{figure}
\begin{center}
\includegraphics[scale=0.6]{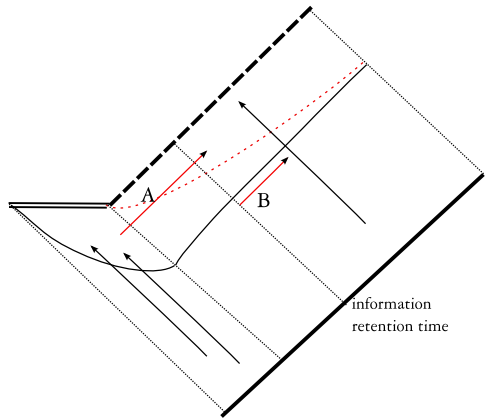}
\caption{\label{fig:complementarity}A schematic diagram for the duplication experiment. An in-falling observer sends his information to the out-going direction (A); his information is also contained in the Hawking radiation (B). If the region beyond the inner horizon is regular, a second observer can see both $A$ and $B$ violating the no cloning theorem.}
\end{center}
\end{figure}

However, if we assume sufficiently large $N$, so that we can extend the causal structure of Figure~\ref{fig:charged} beyond the inner horizon, the duplication experiment will be possible, see Figure~\ref{fig:complementarity}. Since the first observer can send a signal $A$ in the out-going direction in time $\Delta t \sim M$, the signal requires energy $\Delta E \sim 1/M$, which is sufficiently smaller than the black hole mass $M$, and hence this process is possible. Thus, the second observer can see both information $A$ and $B$ in the region beyond the inner horizon, violating black hole complementarity. For further discussions, see \cite{HYZ}.

\section{\label{sec:dis}Conclusion}

We constructed a spherically symmetric charged black hole model with a complex scalar field and a gauge field, and used the $1+1$ dimensional approximation for the renormalized stress tensor to include Hawking radiation.
Solving the equations numerically, we determined the Penrose diagram for dynamical charged black holes (Figure~\ref{fig:charged}).
We found a space-like inner horizon separated from the Cauchy horizon, which is parallel to the out-going null direction. Also, we demonstrated the transition from a charged black hole to a neutral black hole, and observed the generation of a space-like singularity and a second Cauchy horizon.

Inner horizons have often been regarded as Cauchy horizons and null curvature singularities.
However, we demonstrated that, in our model, the inner horizon is distinct from the Cauchy horizon, and it is not a curvature singularity from the general relativistic point of view; it is regular and penetrable. However, if the number of scalar fields are less than $\exp M^{2}$, the curvature functions blow up to greater than the Planckian cutoff beyond the inner horizon, so that we cannot extend physics beyond the inner horizon.
Hence it is fair to say that a charged black hole has a ``physical'' space-like singularity even in dynamical cases (Figure~\ref{fig:charged_reasonable}).
However, if we can assume an exponentially large number of massless fields, we may trust our results beyond the inner horizon. Then we found that strong cosmic censorship and black hole complementarity can be violated.
Of course, the consistency of the assumption of an exponentially large number of massless fields should be checked within string theory.

\ack{The authors would like to thank Evegeny Sorkin, Alex Nielsen,
Chang Sub Shin, Heeseung Zoe, and Yuree S Lim for discussions and
encouragement. They also thank Alexei Starobinsky for useful
historical remarks. This work was supported by Korea Research
Foundation grants (KRF-313-2007-C00164, KRF-341-2007-C00010) funded
by the Korean government (MOEHRD) and BK21.}

\appendix

\section{\label{sec:BasicSchemes}Equations and integration schemes}

We use the gauge-invariant Lagrangian with a complex massless scalar field $\phi$ and an electromagnetic gauge field $A_{\mu}$ \cite{Hawking:1973uf}:
\begin{eqnarray}
\mathcal{L} = - \left(\phi_{;a}+ieA_{a}\phi \right)g^{ab}\left(\overline{\phi}_{;b}-ieA_{b}\overline{\phi}\right)-\frac{1}{8\pi}F_{ab}F^{ab},
\end{eqnarray}
where $F_{ab}=A_{b;a}-A_{a;b}$ and $e$ is the gauge coupling.
From this Lagrangian we can derive the equations of motion for the scalar field and the electromagnetic field:
\begin{eqnarray} \label{scalar_and_Maxwell}
\phi_{;ab}g^{ab}+ieA^{a}\left(2\phi_{;a}+ieA_{a}\phi\right)+ieA_{a;b}g^{ab}\phi &=& 0,
\\
\frac{1}{2\pi}{{F^{b}}_{a}}_{;b}-ie\phi\left(\overline{\phi}_{;a}-ieA_{a}\overline{\phi}\right)+ie\overline{\phi}\left(\phi_{;a}+ieA_{a}\phi\right) &=& 0.
\end{eqnarray}
Also, the energy-momentum tensor becomes
\begin{eqnarray} \label{energy_momentum}
T^{\mathrm{C}}_{ab} &=& \frac{1}{2}\left(\phi_{;a}\overline{\phi}_{;b}+\overline{\phi}_{;a}\phi_{;b}\right)
\nonumber \\
&& {}+\frac{1}{2}\left(-\phi_{;a}ieA_{b}\overline{\phi}+\overline{\phi}_{;b}ieA_{a}\phi+\overline{\phi}_{;a}ieA_{b}\phi-\phi_{;b}ieA_{a}\overline{\phi}\right)
\nonumber \\
&& {}+\frac{1}{4\pi}F_{ac}{F_{b}}^{c}+e^{2}A_{a}A_{b}\phi\overline{\phi}+\frac{1}{2}\mathcal{L}g_{ab}.
\end{eqnarray}
We use the double-null coordinates \cite{Hamade:1995ce},
\begin{eqnarray}
\d{s^{2}} = -\fn{\alpha^{2}}{u,v} \d{u} \d{v} + \fn{r^{2}}{u,v} \d{\Omega^{2}},
\end{eqnarray}
assuming spherical symmetry. Because of spherical symmetry and gauge symmetry, we can choose the gauge field as $A_{\mu}=(a,0,0,0)$ with a single function $a(u,v)$ \cite{OrenPiran}.

Now, we will describe our numerical setup.
We follow the notation of \cite{HodPiran}\cite{SorkinPiran}\cite{OrenPiran}: the metric function $\alpha$, the radial function $r$, the Maxwell field $a$, and the complex massless scalar field $s \equiv \sqrt{4\pi} \phi$, and define
\begin{eqnarray} \label{definition}
h \equiv \frac{\alpha_{u}}{\alpha},\quad d \equiv \frac{\alpha_{v}}{\alpha},\quad f \equiv r_{u},\quad g \equiv r_{v},\quad w \equiv s_{u},\quad z \equiv s_{v}.
\end{eqnarray}
Then the Einstein and energy-momentum tensor components are
\begin{eqnarray} \label{G_and_T}
G_{uu} &=& -\frac{2}{r} (f_{u}-2fh),
 \\
G_{uv} &=& \frac{1}{2r^{2}} \left( 4 rf_{v} + \alpha^{2} + 4fg \right),
 \\
G_{vv} &=& -\frac{2}{r} (g_{v}-2gd),
 \\
G_{\theta\theta} &=& -4\frac{r^{2}}{\alpha^{2}} \left(d_{u}+\frac{f_{v}}{r}\right),
 \\
T^{\mathrm{C}}_{uu} &=& \frac{1}{4\pi} \left[ w\overline{w} + iea(\overline{w}s-w\overline{s}) +e^{2}a^{2}s\overline{s} \right],
 \\
T^{\mathrm{C}}_{uv} &=& \frac{{a_{v}}^{2}}{4\pi\alpha^{2}},
 \\
T^{\mathrm{C}}_{vv} &=& \frac{1}{4\pi} z\overline{z},
 \\
T^{\mathrm{C}}_{\theta\theta} &=& \frac{r^{2}}{4\pi\alpha^{2}} \left[ (w\overline{z}+z\overline{w}) + iea(\overline{z}s-z\overline{s})+\frac{2{a_{v}}^{2}}{\alpha^{2}} \right].
\end{eqnarray}
The scalar field and Maxwell field equations become
\begin{eqnarray} \label{scalar_and_Maxwell_2}
rz_{u}+fz+gw+iearz+ieags+ies\frac{\alpha^{2}q}{4r}&=&0,
\\
\left( \frac{r^{2}a_{v}}{\alpha^{2}} \right)_{v}+\frac{ier^{2}}{4}(z\overline{s}-s\overline{z})&=&0,
\end{eqnarray}
where
\begin{eqnarray}
q \equiv \frac{2r^{2}a_{v}}{\alpha^{2}}
\end{eqnarray}
can be interpreted as the electric charge.

We use the semi-classical Einstein equation,
\begin{eqnarray}
G_{\mu\nu}=8\pi \left( T^{\mathrm{C}}_{\mu\nu}+\langle \hat{T}^{\mathrm{H}}_{\mu\nu} \rangle \right)
\end{eqnarray}
to include Hawking radiation.
Spherical symmetry makes it is reasonable to use the $1+1$-dimensional results for $\langle \hat{T}^{\mathrm{H}}_{\mu\nu} \rangle$\cite{Davies:1976ei} divided by $4\pi r^{2}$ \cite{SorkinPiran}:
\begin{eqnarray}
\langle \hat{T}^{\mathrm{H}}_{uu} \rangle &=& \frac{P}{4\pi r^{2}}\left(h_{u}-h^{2}\right),
 \\
\langle \hat{T}^{\mathrm{H}}_{uv} \rangle = \langle \hat{T}^{\mathrm{H}}_{vu} \rangle &=& -\frac{P}{4\pi r^{2}}d_{u},
 \\
\langle \hat{T}^{\mathrm{H}}_{vv} \rangle &=& \frac{P}{4\pi r^{2}}\left(d_{v}-d^{2}\right),
\end{eqnarray}
with $P \equiv Nl_{\mathrm{Pl}}^2 / 12\pi$, where $N$ is the number of massless scalar fields and $l_{\mathrm{Pl}}$ is the Planck length.

Finally, we can list the simulation equations:
\begin{enumerate}
\item \textit{Einstein equations:}
\begin{eqnarray} \label{E1}
d_{u} = h_{v} &=& \frac{1}{(1-\frac{P}{r^{2}})} [ \frac{fg}{r^{2}} + \frac{\alpha^2}{4r^{2}} - \frac{\alpha^{2} q^{2}}{2 r^{4}} - \frac{1}{2}(w\overline{z}+\overline{w}z)\\ \nonumber
& & - \frac{iea}{2}(s\overline{z}-\overline{s}z) ], \\
\label{E2}
g_{v} &=& 2dg - rz\overline{z} - \frac{P}{r}(d_{v}-d^{2}),\\
\label{E3}
f_{u} &=& 2fh - rw\overline{w} - iear(\overline{w}s-w\overline{s}) - e^{2}a^{2}rs\overline{s} \nonumber \\
& & - \frac{P}{r}(h_{u}-h^{2}),\\
\label{E4} f_{v} = g_{u} &=& -\frac{fg}{r} -
\frac{\alpha^{2}}{4r} + \frac{\alpha^{2} q^{2}}{4 r^{3}} -
\frac{P}{r} d_{u}.
\end{eqnarray}
\item \textit{Maxwell equations:}
\begin{eqnarray} \label{M1}
a_{v} &=& \frac{\alpha ^{2} q}{2 r^{2}}, \\
 \label{M2}
q_{v} &=& -\frac{ier^{2}}{2} (\overline{s}z-s\overline{z}).
\end{eqnarray}
\item \textit{Scalar field equations:}
\begin{eqnarray} \label{S1}
z_{u} = w_{v} = - \frac{fz}{r} - \frac{gw}{r} - \frac{iearz}{r} - \frac{ieags}{r} - \frac{ie}{4r^{2}}\alpha^{2}qs.
\end{eqnarray}
\end{enumerate}
Also, we have the definitions in Equation (\ref{definition}).
If we substitute Equation (\ref{E1}) into Equation (\ref{E4}), then all equations contain only one derivative, except Equation (\ref{E2}) and (\ref{E3}).

We can choose two different integration schemes corresponding to a choice of the equation. First, we can get $r$ from $g$ using Equation (\ref{E2}), $\alpha$ from $d$, and $s$ from $z$. Second, we can get $r$ from $f$ using Equation (\ref{E3}), $\alpha$ from $h$, and $s$ from $w$. We mainly used the first integration scheme. If we call the radial function of the first scheme $r_{(v)}$ and the radial function of the second scheme $r_{(u)}$, then $r_{(v)}$ and $r_{(u)}$ should be the same. We compared them to check the consistency of our simulation.

We solved these equations using the second order Runge-Kutta method \cite{nr} following \cite{OrenPiran}.
Equation (\ref{E1}) has a singularity at $r=\sqrt{P}$ and we may regard it as the central singularity due to the semi-classical approximation. For consistency $\sqrt{P}$ should be sufficiently smaller than the size of the black hole in the simulation.

\section{\label{sec:initialCond}Initial conditions}

We need initial conditions for all functions ($\alpha, r, g, f, h, d, s, w, z, a, q$) on the initial $u=u_{\mathrm{i}}$ and $v=v_{\mathrm{i}}$ surfaces, where we set $u_{\mathrm{i}}=v_{\mathrm{i}}=0$.

We have gauge freedom to choose the initial $r$ function. Although all constant $u$ and $v$ lines are null, there remains freedom to choose the distances between these null lines. Here, we choose $r(0,0)=r_{0}$, $f(u,0)=r_{u0}$, and $g(0,v)=r_{v0}$, where $r_{u0}<0$ and $r_{v0}>0$ so that the radial function for an in-going observer decreases and for an out-going observer increases.

We use a shell-shaped scalar field, and hence its inside is not affected by the shell.
Thus, we can simply choose $q(u,0)=a(u,0)=0$ and $\alpha(u,0)=1$. Also, $s(u,0)=w(u,0)=h(u,0)=0$ holds.
Then, since the mass function, defined in Equation (\ref{mass}), should vanish at $u=v=0$, it is convenient to choose $r_{u0}=-1/2$ and $r_{v0}=1/2$.

We need more information to determine $d, g$, and $z$ on the $v=0$ surface. We get $d$ from Equation (\ref{E1}), $g$ from Equation (\ref{E4}), and $z$ from Equation (\ref{S1}).

We can choose an arbitrary function for $s(0,v)$. For example, to make a black hole, we use
\begin{eqnarray}
s(0,v)=A \sin ^{2} \left( \pi \frac{v}{v_{\mathrm{f}}} \right) \exp \left( \pm 2 \pi i \frac{v}{v_{\mathrm{f}}} \right)
\end{eqnarray}
for $0\leq v \leq v_{\mathrm{f}}$ and otherwise $s(0,v)=0$, where $v_{\mathrm{f}}$ is the width of the pulse and $A$ is the amplitude.
Then we obtain $z(0,v)$.
Note that, according to Equation (\ref{M2}), if we represent the scalar field as $s=|s| \exp(i \Omega)$, then $q_{v}=er^{2} |s|^{2} \Omega_{v}$ holds, and the sign of the exponent determines the sign of charge.

Also from Equation (\ref{E2}), we can use $d = rz\overline{z}/2g$ on the $u=0$ surface, since we can assume that there is no Hawking effect on the initial surface, and we get $d(0,v)$. By integrating $d$ along $v$, we get $\alpha(0,v)$.

We need more information for $h, f, w, a$, and $q$ on the $u=0$ surface. We obtain $q$ from Equation (\ref{M2}) and $a$ from Equation (\ref{M1}). Then, we get $h$ from Equation (\ref{E1}), $f$ from Equation (\ref{E4}), and $w$ from Equation (\ref{S1}). This finishes the assignments of initial conditions.

We choose $r_{0}=10$ and $v_{\mathrm{f}}=20$, leaving the three free parameters $(e, P, A)$.

\section{\label{sec:consistency}Convergence and consistency checks}

\begin{figure}
\begin{center}
\includegraphics[scale=0.95]{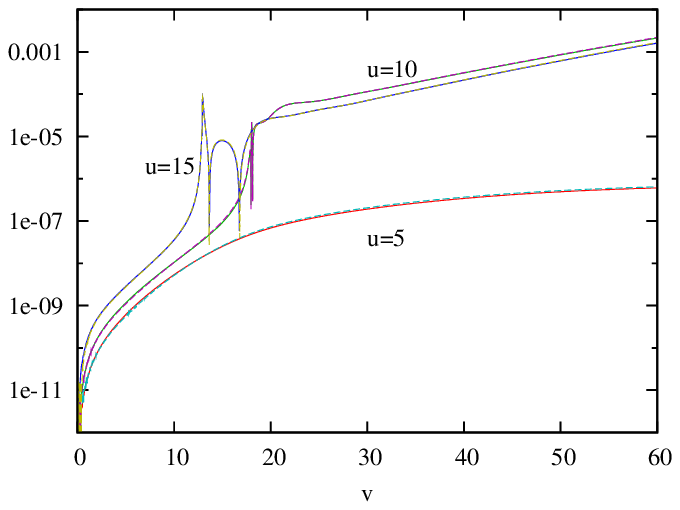}
\includegraphics[scale=0.95]{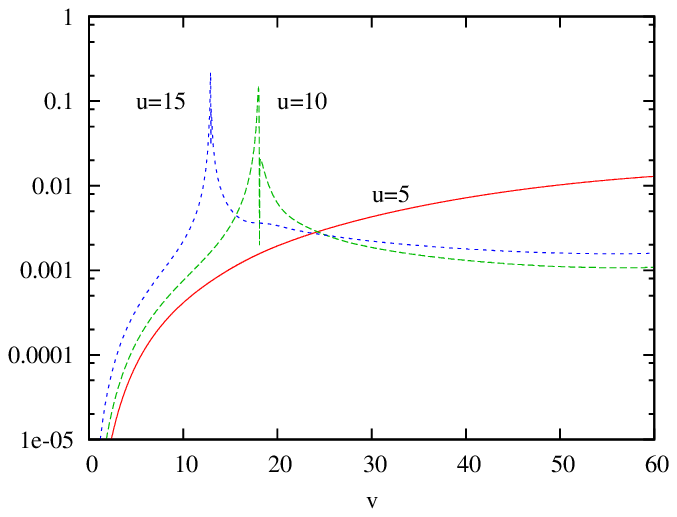}
\caption{Plots of errors with different step sizes. Here, we plot $|r_{(1)}-r_{(2)}|/r_{(2)}$ (solid curves) and $4|r_{(2)}-r_{(4)}|/r_{(4)}$ (dashed curves) along a few constant $u$ lines, where $r_{(n)}$ is calculated in an $n$ times finer simulation than $r_{(1)}$. This shows that our simulation converges to second order. \label{fig:convergence}}
\caption{$|r_{(v)}-r_{(u)}|/r_{(v)}$ along a few constant $u$ lines, where $r_{(v)}$ is calculated by integrating $g$ using Equation (\ref{E2}), and $r_{(u)}$ is calculated by integrating $f$ using Equation (\ref{E3}). \label{fig:consistency}}
\end{center}
\end{figure}

To check the convergence of our simulations, we compared the results for $(e=0.1,P=0.1,A=0.25)$ using different step sizes: 1, 2, and 4 times finer.
In Figure~\ref{fig:convergence}, we see that the difference between 1 and 2 times finer is 4 times the difference between 2 and 4 times finer, and thus our simulation converges to second order with errors $\lesssim 0.1\%$.

To check the consistency, we used two independent integration schemes as mentioned in \ref{sec:BasicSchemes}.
Figure~\ref{fig:consistency} shows that the difference between the two schemes is $\lesssim 1 \%$.

Near the outer horizon, the radial function $r$ changes rapidly causing the slow convergence shown by the spikes in Figure~\ref{fig:convergence} and \ref{fig:consistency}.
To avoid this problem, we chose the step size $\Delta u$ adaptively so that the ratio $r_{u}\Delta u / r$ was constant \cite{OrenPiran}.

\begin{figure}
\begin{center}
\includegraphics[scale=0.3]{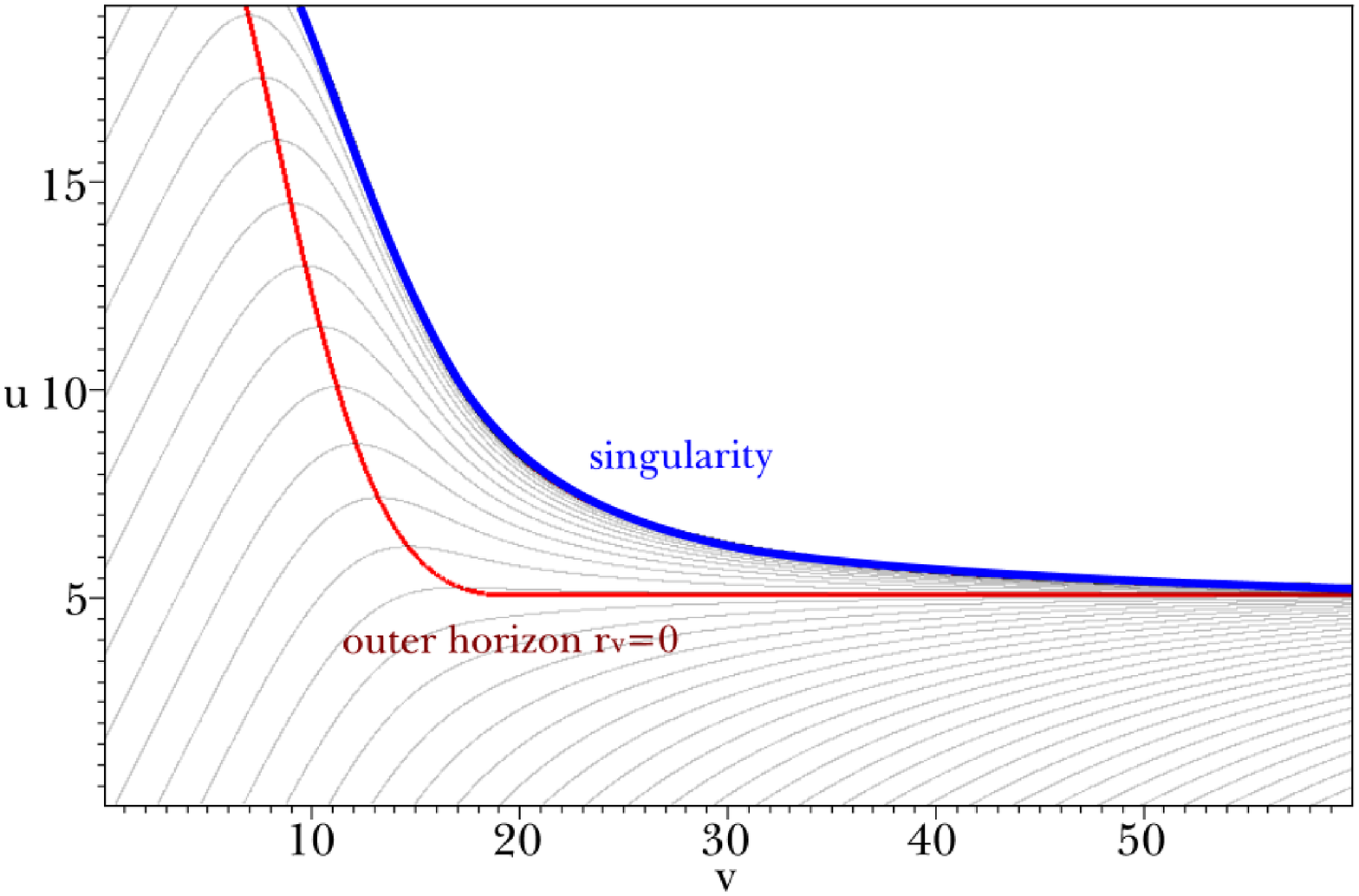}
\includegraphics[scale=0.3]{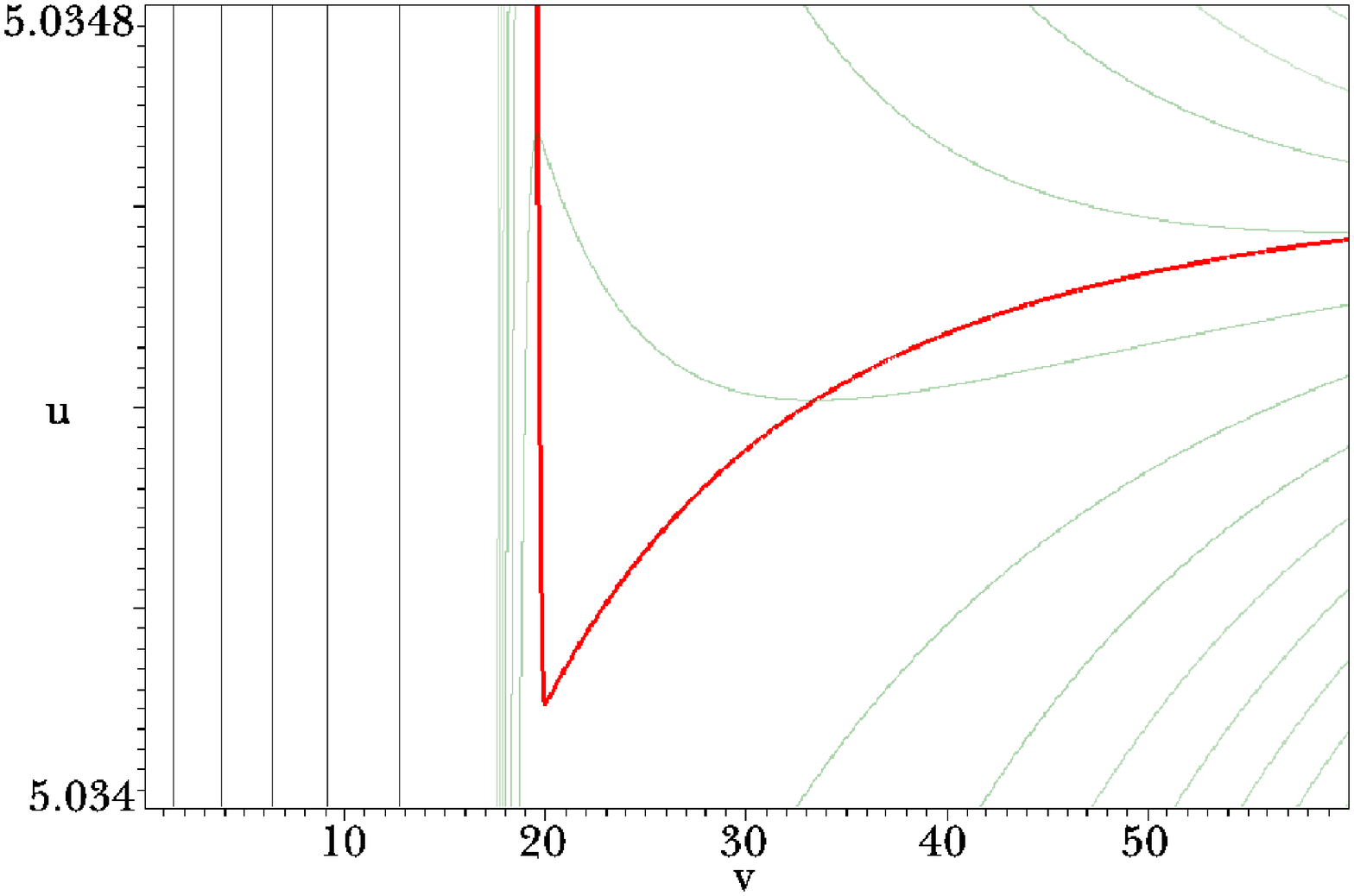}
\caption{Contour diagram of $r$ for a neutral black hole $(e=0,P=0.1,A=0.25)$. One can see a space-like singularity. \label{fig:Schwarzschild_overall}}
\caption{Detailed plot near the outer horizon of Figure~\ref{fig:Schwarzschild_overall}. The outer horizon ($r_{v}=0$, red curve) bends in a time-like direction as the matter supply ends. Here, spacing is $1$ for black contours and $0.002$ for green contours. \label{fig:Schwarzschild}}
\end{center}
\end{figure}

\begin{figure}
\begin{center}
\includegraphics[scale=0.55]{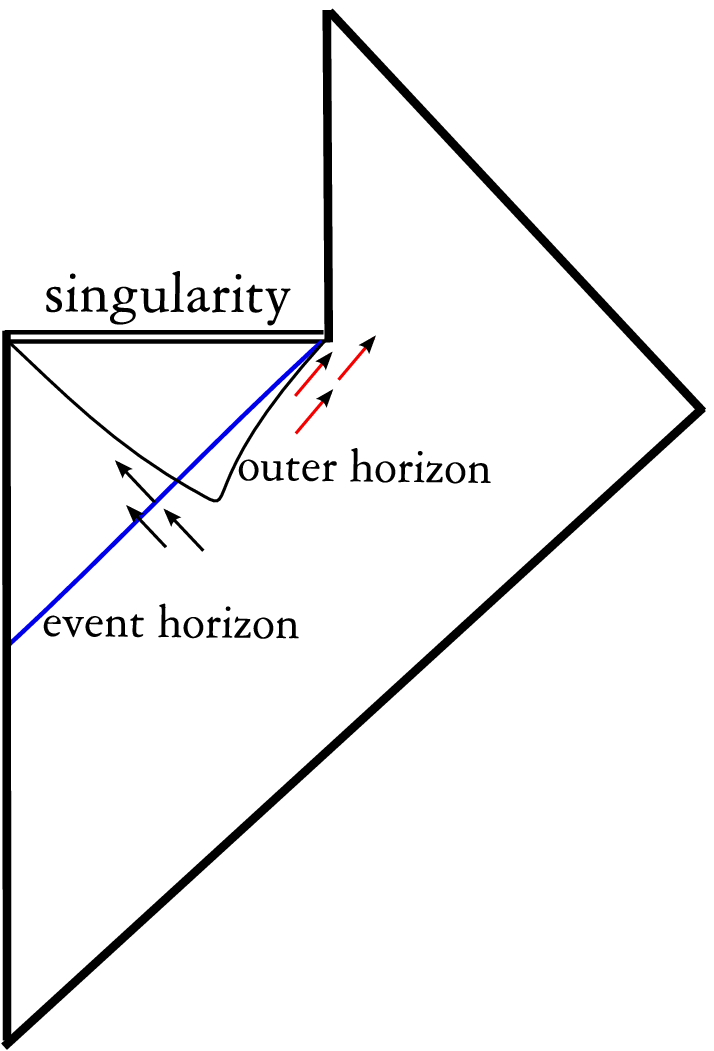}
\caption{\label{fig:Schwarzschild_theory}The causal structure of a neutral black hole \cite{inforpara}.}
\end{center}
\end{figure}

Finally, we checked whether our simulation gives the physically correct picture for a neutral black hole \cite{SorkinPiran}.
Figure~\ref{fig:Schwarzschild_overall} and \ref{fig:Schwarzschild} are contour diagrams for $r$ and the outer horizon $r_{v} = 0$ for $(e=0, P=0.1, A=0.25)$.
As the matter collapses, the outer horizon bends in a space-like direction. After the matter collapse ends, the outer horizon bends in a time-like direction due to the Hawking radiation.
These phenomena are consistent with the properties of local horizons \cite{local_horizon}.
Moreover, one can see the space-like central singularity at $r=\sqrt{P}$.
This confirms that our simulation gives the correct result for a neutral black hole (Figure~\ref{fig:Schwarzschild_theory}) \cite{CGHS}.


\end{document}